\documentclass[fleqn,usenatbib]{mnras}

\usepackage{newtxtext,newtxmath}

\usepackage[T1]{fontenc}

\DeclareRobustCommand{\VAN}[3]{#2}
\let\VANthebibliography\thebibliography
\def\thebibliography{\DeclareRobustCommand{\VAN}[3]{##3}\VANthebibliography}

\usepackage{graphicx}	
\usepackage{amsmath}	
\usepackage{bm}

\title[Measuring the matter fluctuations in the LU]{Measuring the matter fluctuations in the Local Universe with the ALFALFA catalogue}

\author[C. Franco et al.]{
Camila Franco,$^{1}$\thanks{E-mail: camilafranco@on.br}
Jezebel Oliveira,$^{2}$
Maria Lopes$^{1}$
Felipe Avila$^{1}$
and Armando Bernui$^{1}$
\\
$^{1}$Observatório Nacional, Rua General José Cristino, 77, São Cristóvão, 20921-400, Rio de Janeiro, RJ, Brazil\\
$^{2}$Observatório do Valongo, Ladeira do Pedro Antônio, 43, Centro, 20080-090, Rio de Janeiro, RJ, Brazil\\
}

\date{Accepted 2025 January 10. Received 2024 November 8; in original form 2024 September 19}

\pubyear{\the\year{}}

\begin{document}
\label{firstpage}
\pagerange{\pageref{firstpage}--\pageref{lastpage}}
\maketitle

\begin{abstract}
The standard model of cosmology describes the matter fluctuations through the matter power spectrum, where $\sigma_{8} \equiv \sigma_{8,0} \equiv \sigma_{8}(z = 0)$, defined at the scale of $8\,h^{-1}$ Mpc, acts as a normalisation parameter. Currently, the literature reports measurements of $\sigma_{8}$ analysing different cosmic tracers, where some of these results were obtained assuming a fiducial cosmology. In this study we measure, in a model-independent approach, the matter fluctuations in the Local Universe using HI extragalactic sources mapped by the ALFALFA survey. Our analyses allow us to test the standard cosmological model under extreme conditions in the highly non-linear Local Universe, quantifying the amplitude of the matter fluctuations there. 
Our work directly measures $\sigma_{8}$ using the 3-dimensional distances of the HI sources determined by the ALFALFA survey without assuming a fiducial cosmology, resulting in a robust model-independent measurement of $\sigma_{8}$. 
Our methodology involves the construction of suitable mock catalogues to simulate the large scale structure features observed in the data, applying the 2-point correlation function, and making use of Markov Chain Monte Carlo methods to estimate the parameters. 
Analysing these data we measure $\sigma_8 = 0.78 \pm 0.04$ for $h = 0.6727$, $\sigma_8 = 0.80 \pm 0.05$ for $h = 0.698$, and $\sigma_8 = 0.83 \pm 0.05$ for $h = 0.7304$. Considering the data pairs $(\sigma_8, H_0)$ from the Planck CMB and Atacama Cosmology Telescope (ACT) CMB-lensing analyses, our measurement agrees with them within $1\,\sigma$ confidence level. From a model-independent perspective, we find that the scale where the matter fluctuation is $1$ is $R = 7.2 \pm 1.5~\text{Mpc}$.
\end{abstract}



\section{Introduction}\label{sec:introduction}
Cosmic structures form by gravitational instability, when matter contained in a sufficiently dense region collapses. 
The standard cosmological model reproduces quantitatively the main features of the evolution of matter fluctuations through the matter power spectrum. 
The normalisation parameter of this spectrum, $\sigma_{8} \equiv \sigma_{8,0}$, is defined as the amplitude of the matter fluctuations at redshift $z=0$, at the scale of $8\,h^{-1}$ Mpc~\citep{Peebles67,FKP1994,T_Padmanabhan,Mo}.

A current tension in the value of $\sigma_{8}$ is reported in the literature due to the different values obtained using cosmic microwave background (CMB) and large scale structure (LSS) data~\citep{Nunes21,Abdalla22}. 
Notice that the Planck CMB value, $\sigma_{8}^{\text{Pla}}$, is not a direct measurement but a derived one, 
assuming the $\Lambda$CDM model to describe the evolution of the primordial perturbations, and 
using Bayesian analyses to best-fit the CMB temperature fluctuations and combining these data with other cosmic probes. 
As well, the literature reports $\sigma_{8}$ values obtained through statistical analyses that combine cosmological probes using, sometimes, a model-dependent approach to obtain 
$\sigma_{8}$~\citep{Planck20,Heymans21,Avila2022,ACT2024}.

The cosmological parameter $\sigma_8$, that quantifies the matter 
fluctuation, is related to $\sigma_8^{\text{tr}}$, the fluctuation of the cosmic tracer, through a linear bias; $\sigma_8$ is used to normalise the matter power spectrum, therefore it is a very important parameter in modern 
cosmology~\citep{kaiser88,Juszkiewicz10}. 
In general, measurements of $\sigma_8$ in redshift surveys use cosmic objects that are reliable matter tracers, i.e., $b \sim 1$. 
By measuring the variance of the number density of galaxies, one effectively obtains the variance of matter, enabling a comparison with diverse measurements reported in the literature~\citep{Borgani95}. 
The fluctuation in the number density of galaxies in the Local Universe is roughly of order 1 on scales of $8\,h^{-1}$\,Mpc 
($h$ is defined by $H_0 = 100\,h$ km s$^{-1}$ Mpc$^{-1}$). 

Using Counts-in-Cells (CIC) in the IRAS survey, \cite{Efstathiou90} found $\sigma^2 = 0.26$ at the effective distance $R = 30h^{-1}$ Mpc \,and\, $\sigma^2 = 0.21$ at $R = 40h^{-1}$~Mpc, in disagreement with the prevailing standard cosmological model of that time, i.e., the Cold Dark Matter (CDM) with $\Omega = 1$, suggesting that the analysed data exhibited more clustering than predicted by the model. 
In a later study, \cite{Efstathiou95} revisited this technique with two surveys of IRAS galaxies and found a better agreement with the CDM model. 
However, taken into account the error interval, other models were also conceivable, such as those with lower values of the matter density parameter, such as the $\Lambda$CDM model. 
Simultaneously, \cite{Borgani95} conducted an extensive work to validate simulated catalogues in the study of cluster distribution, employing various statistical methods to measure $\sigma_8$. 
The application of this methodology to the Abell/ACO cluster redshift sample showed a preference for the $\Lambda$CDM model in explaining Abell/ACO data clustering. 
More recently, \cite{Repp20} applying the CIC approach in the SDSS Main Galaxy Sample, constrained $b$ and $\sigma_8$ simultaneously, obtaining values of $b = 1.36 \pm 0.14$ and $\sigma_8 = 0.94 \pm 0.11$, which agrees well with previous findings. 
Overall, these studies indicate that at scales close to or above 
$8\,h^{-1}$~Mpc, the clustering of galaxies is in excellent agreement with linear perturbation theory, provided that the bias is approximately~1.

The aim of this study is to analyse the mass fluctuations within the Local Universe across several scales, for this we use as a cosmic tracer the HI extragalactic sources mapped by the ALFALFA survey. 
One of the main motivations to perform this study is to test the standard cosmological model at extremely low redshift, which means that we are dealing with non-linear scales. 
In fact, it is well known that the Local Universe exhibits significant non-linear fluctuations in structure growth, providing a unique opportunity to assess the limit of linear theory across various scales~\citep{Borgani95,
Martin12,Papastergis13,Avila18,Avila19,Avila21,Franco24}; 
for this task we use cosmological distances in real space. 
Notice that the ALFALFA catalogue provides the distance values of the observed HI sources, where these quantities are not directly measured, instead they are estimated assuming a semi-analytical velocity field model that uses a default value for $H_0 = 70$ km s$^{-1}$ Mpc$^{-1}$~\citep{Masters2005}. 
This suggests the possibility that the distances obtained with this flow model could be biased, and our $\sigma_8$ measurement too. 
To this end, as explained below, we perform consistency analyses to investigate the impact of assuming this specific value for $H_0$ in our measurement of $\sigma_8$~\citep{Jones18}.

In addition, because the HI sources are effective tracers of dark matter for scales below $10\,h^{-1}$~Mpc~\citep{Martin12,Papastergis13,Jones16}, 
studying the ALFALFA catalogue one can obtain valuable insights into the matter fluctuations in the Local Universe. 
Observe that $\sigma_8$ is an important cosmological observable that helps to understand the role of dark matter in the dynamics of cosmic structures growth~\citep{Avila2022}.
Our analyses differs from others in the literature in that: 
(i) we perform a direct measurement of $\sigma_{8}$ using the ALFALFA survey data in the Local Universe, $z \approx 0$; 
(ii) we measure $\sigma_{8}$ for HI extragalactic sources, which are good tracers of matter because $b \sim 1$~\citep{Obuljen19}; 
(iii) we use the 3-dimensional (3D) distances provided by the ALFALFA catalogue, therefore our analysis is model independent; 
(iv) we perform diverse robustness tests to confirm the validity of our results.

Our work is structured as follows: the ALFALFA catalogue is presented in section~\ref{sec:data}, which details the process of selecting objects for analyses, as well as the construction of the random catalogues and of the covariance matrix, essential components for our analyses. 
In section~\ref{sec:methodology}, we elaborate on the methodology used to calculate the matter fluctuations, $\sigma(R)$, and its corresponding error. The results of our analyses and our conclusions are presented in sections \ref{sec:results} and \ref{sec:conclusions}, respectively. 
All robustness tests that support our main results are presented in the appendix.

\section{The ALFALFA catalogue}\label{sec:data}

The ALFALFA survey was initiated in 2005 with the scientific aim of obtaining measurements of the mass and luminosity functions through extragalactic sources in the HI 21 cm line. 
By its conclusion in 2018, it covered an area in the sky of around $6900 \deg^2$ at redshift $z < 0.06$. 
At the end, $31502$ HI sources were observed. Due to the limitations of the radio telescope used to detect the HI sources, the observations were divided between the northern hemisphere ($7^{h} 20^{m} <$ RA $< 3^{h} 15^{m}$) and the southern hemisphere ($7^{h} 20^{m} <$ RA $< 21^{h} 30^{m}$). 

The ALFALFA data show a significant distinction based on the signal-to-noise ratio, categorising them by specific codes: Code 1: involves observations with high signal-to-noise ratio, confirmed in the optical part; Code 2: refers to sources with low signal-to-noise ratio, confirmed in the optical part but considered unreliable; Code 9: relates to high-velocity HI clouds without optical confirmation. In this work, we exclusively use Code 1 sources, following \cite{Haynes18} 
(for large-scale structure analyses using the ALFALFA survey see, e.g.,~\cite{Avila18,Avila21,Franco24}).

\subsection{Data selection}\label{sec:data-selection}

The ALFALFA collaboration estimates distances of the HI extragalactic sources following two approaches, depending if $cz_{\odot}$ is less than or greater than $6000$ km s$^{-1}$~\citep{Haynes18}. 
Because we are interested in studying matter clustering in the Local Universe, we select for analysis the HI sources with $cz_{\odot} < 6000$ km s$^{-1}$, corresponding to distances $\lesssim 85$ Mpc. 
Moreover, to avoid highly non-linear effects, we consider cosmic objects at distances greater than $20$ Mpc. 
That is, we select for our analysis $3682$ HI extragalactic sources of the Local Universe with distances in the range $20-85$ Mpc and with $cz_{\odot} < 6000$ km s$^{-1}$~\citep{Avila21,Avila23}. 

\begin{figure}
\begin{minipage}[b]{\linewidth}
\centering
\includegraphics[width=\textwidth]{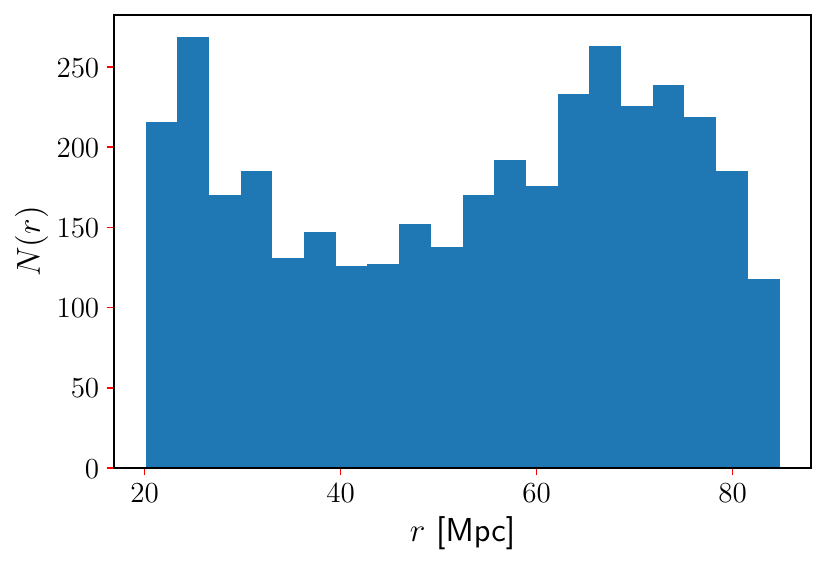}
\end{minipage}
\caption{Distances distribution of the selected sample, with $3682$ HI extragalactic sources.}
\label{fig:histogram}
\end{figure}

\begin{figure*}
    \begin{minipage}[b]{\linewidth}
    \centering
    \includegraphics[scale=0.8]{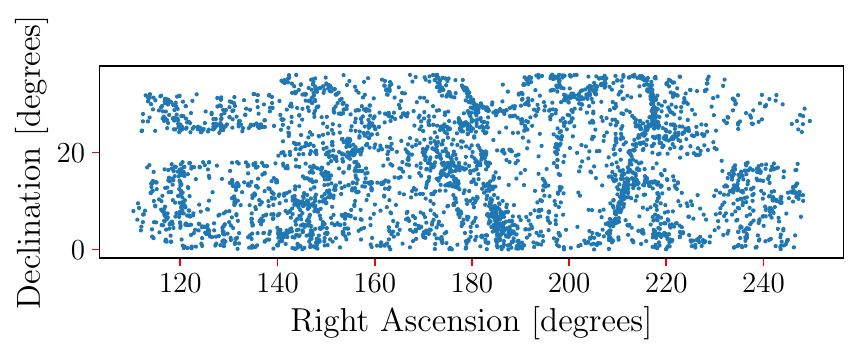}
    \end{minipage}
    \caption{Footprint of the selected sample in equatorial coordinates, localised in the Northern Hemisphere.}
    \label{fig:footprint}
\end{figure*}

\begin{figure}
    \begin{minipage}[b]{\linewidth}
    \centering
    \includegraphics[scale=0.8]{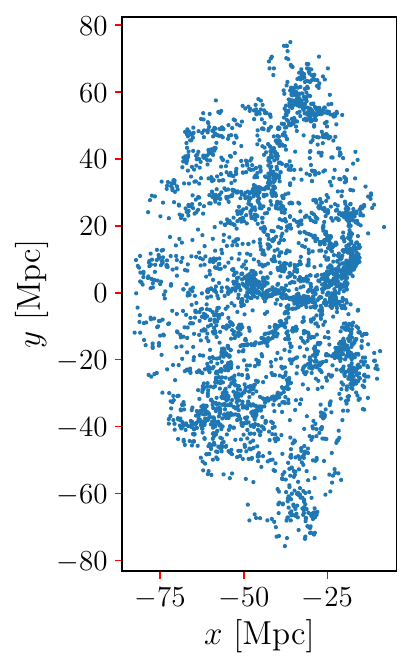}
    \end{minipage}
    \caption{Cartesian projection of the selected sample.}
    \label{fig:cartesian}
\end{figure}

The survey was subdivided into observations conducted in the Northern and Southern hemispheres. 
In this work we analyse the region in the Northern hemisphere, 
which covers a larger area and encompasses a greater number of detected HI sources. 
Figure~\ref{fig:histogram} shows the distribution of this sample, 
while figures~\ref{fig:footprint} and~\ref{fig:cartesian} illustrate the footprint and the Cartesian projection of the selected sample, respectively.

One potential problem in our analyses arises from the fact that distances in the ALFALFA catalogue were not directly measured by the survey but estimated using a semi-analytical flow model
that assumes the value $H_0 = 70$ km s$^{-1}$ Mpc$^{-1}$ for the Hubble constant~\citep{Masters2005}. 
To investigate a possible bias of this particular choice of $H_0$ in our analyses we produce a number of simulated distance catalogues by randomly sampling the value of the Hubble constant from a Gaussian distribution; 
then we analyse these simulated data according to our methodology looking for a possible bias in our $\sigma_8$ measurement. We perform this study in the Appendix~\ref{AppendixC}.

\subsection{Random catalogues}\label{sec:random-catalog}
The construction of random catalogues, i.e., a point distribution with the same observational features as the data but without correlations between the objects, is an essential step in obtaining the correlation function, $\xi(r)$. 
The number density of this catalogue is tens of times greater than the sample data in order to reduce the statistical noise in the correlation function~\citep{Keihanen19,Avila24}. 

\begin{figure*}
    \begin{minipage}[b]{\linewidth}
    \centering
    \includegraphics[scale=0.8]{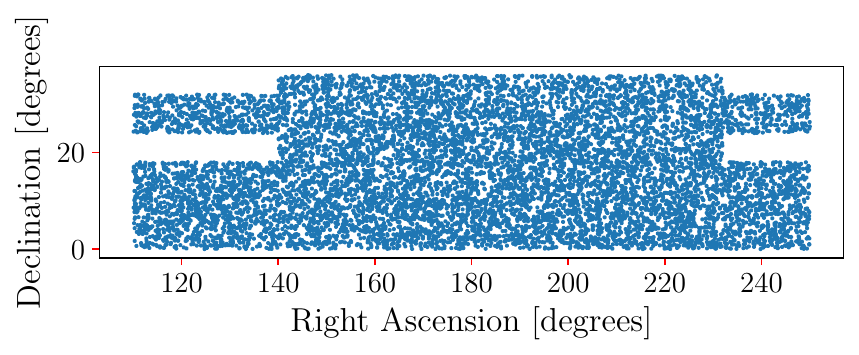}
    \end{minipage}
    \caption{Footprint of the random sample.}
    \label{fig:random_footprint}
\end{figure*}

\begin{figure}
    \begin{minipage}[b]{\linewidth}
    \centering
    \includegraphics[scale=0.8]{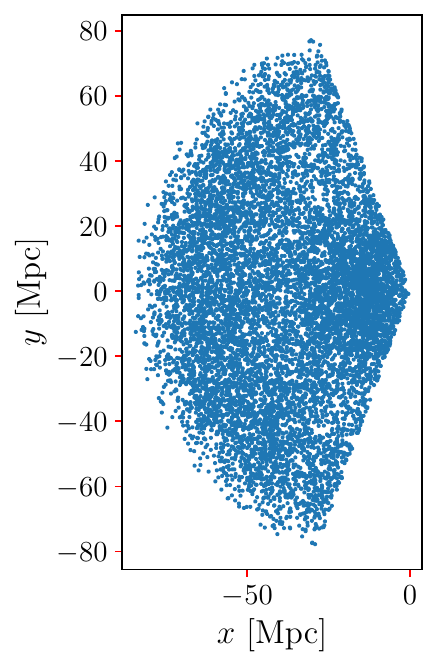}
    \end{minipage}
    \caption{Cartesian projection of the random sample with the footprint of the sample in analysis (see figure~\ref{fig:cartesian}).}
    \label{fig:random_cartesian}
\end{figure}

For this work, we implemented the publicly available code \textsc{randomsdss}\footnote{\url{https://github.com/mchalela/RandomSDSS}}\citep{Chalela_RandomSDSS_2021}, which allows us to generate a random redshift distribution from the original measurements while keeping the selection features unchanged. The angular distribution is generated following the algorithm from \cite{Franco24}. 
The random catalogue has ten times as many points as the data sample. 
Figures \ref{fig:random_footprint} and \ref{fig:random_cartesian} show the angular coordinates and the Cartesian projection of the random catalogue, respectively.

\subsection{Mock catalogues}\label{sec:mock-catalog}

The large-scale structure of the Universe is better understood 
knowing the correlation between its components, information that is 
encoded in the covariance of the 2-point correlation function of cosmic objects, like the galaxies. 
In fact, this function encodes crucial information regarding a galaxy survey, including the survey volume, masking effects, and the average galaxy density (see, for instance, ref.~\cite{Santi22}).
Calculating the variance of the correlation function needs the evaluation of high-order correlations, however, accurate and accessible methods have been created to estimate it. \cite{Martinez01} showed six of such methods, made for specific estimators. 
These methods usually use either theoretical ideas or the observed galaxy distribution itself to calculate the error, like jackknife and bootstrap techniques~\citep{Norberg09}. 
Nevertheless, the best method found by~\cite{Martinez01} uses artificial galaxy distributions, either with N-body simulations or random methods with a known correlation function; it works better because one can obtain the dispersion from multiple correlation function measurements calculated in a way similar to that observed in the sample under analysis. 

In this study, we adopt a stochastic model known as Cox processes~\citep{Martinez98,Pons99,Gausmann2024}. 
While this model does not encompass all the non-linear physical processes involved in structure formation, as seen in N-body simulations~\citep{Pandey10}, Cox processes are advantageous due to their simplicity, low computational requirements, and the availability of their analytic 2-point correlation function~\citep{Martinez01,Avila18}, 
\begin{equation}\label{eq:Cox_equation}
\xi_{\text{Cox}}(r) = \frac{1}{2\pi r^2 L_V} - \frac{1}{2\pi r\,l\,L_V} \,,
\end{equation}
where $l$ and $L_V$ are parameters of the point process generation. 
These mock catalogues are constructed in the following way: we distribute $n_s$ segments of length $l$ in a cube of side $L$. The average number of segments per unit volume, $\lambda_s$, times the length $l$ defines the density of segments, $L_{V}=\lambda_s l$. 
The number of points on each segment are on average the same. 
Let $\lambda_l$ be the number of points per unit length, 
then the intensity of the point process is $\lambda=\lambda_l L_V$. 

Due to the methodology of our analyses, we seek that the correlation function obtained from the Cox processes be close to the correlation function of the HI sources. 
The dispersion of the $n$ realisations, described above, will 
represent the error of the correlation function of the sample of 
HI sources. 
Thus, the parameters of the Cox process are adjusted in such a way 
as to reproduce the observational features presented by the sample 
in analysis. 
However, an initial step can be taken by assuming the Cosmological Principle: on sufficiently large scales the Universe is statistically homogeneous and isotropic. This imposes restrictions on the values of $l$. 
It can be shown that the scale of homogeneity in the Cox process is $R=4\,l$. 
Assuming a minimum scale for homogeneity of 100 Mpc, one expects $l>25$ Mpc. 
Our procedure starts fixing $l=25$ and then chose $L_{V}$ when getting a number density similar to our HI sample.

\begin{figure*}
    \begin{minipage}[b]{\linewidth}
    \centering
    \includegraphics[scale=0.8]{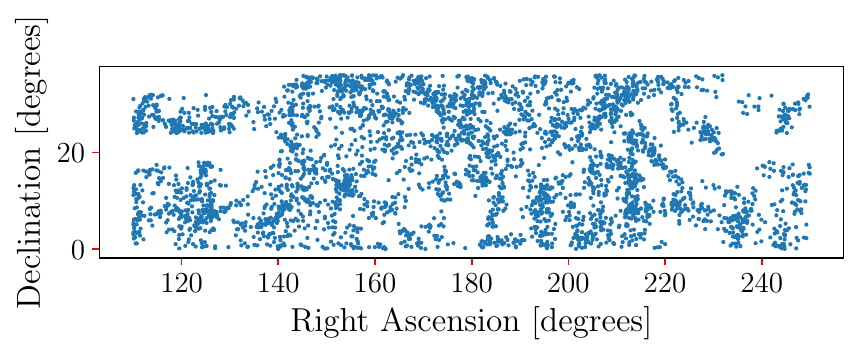}
    \end{minipage}
    \caption{Footprint of a mock realisation.}
    \label{fig:mock_footprint}
\end{figure*}

\begin{figure}
    \begin{minipage}[b]{\linewidth}
    \centering
    \includegraphics[scale=0.8]{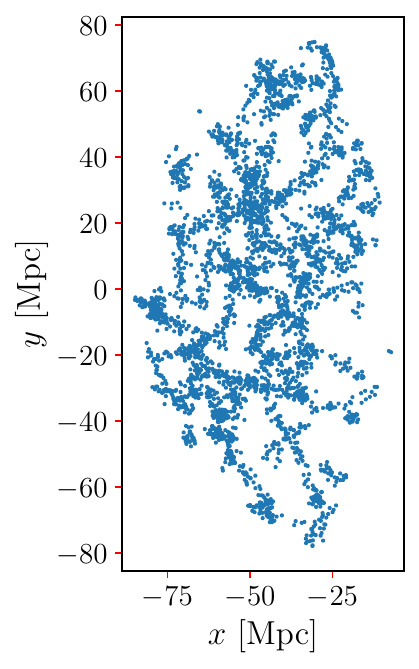}
    \end{minipage}
    \caption{Cartesian projection of a mock realisation.}
    \label{fig:mock_cartesian}
\end{figure}

For this work, we set the following values for the Cox process parameters: 
$\{l, L_{V}, \lambda_l\}=\{25, 0.008, 1.8\}$. 
To obtain a representative dispersion, we construct 1000 mocks. 
In the figures \ref{fig:mock_footprint} and \ref{fig:mock_cartesian}, we show an example of a realisation of a Cox process 
in angular coordinates and its Cartesian projection, respectively. 
Visually, a realisation from the Cox process appears more clustered on small scales. 
This is because, in order to obtain good agreement on the scales analysed, it was necessary to increase the average number of points on the randomly distributed segments. 
As we are not analysing a large cosmological volume, we do not observe the correlation function of the HI sources going to zero, and consequently it is difficult to correctly adjust simultaneously all the scales of the Cox process, as $l$ indicates the scale where the equation~(\ref{eq:Cox_equation}) is zero. 

\begin{figure}
    \begin{minipage}[b]{\linewidth}
    \centering
    \includegraphics[width=\textwidth]{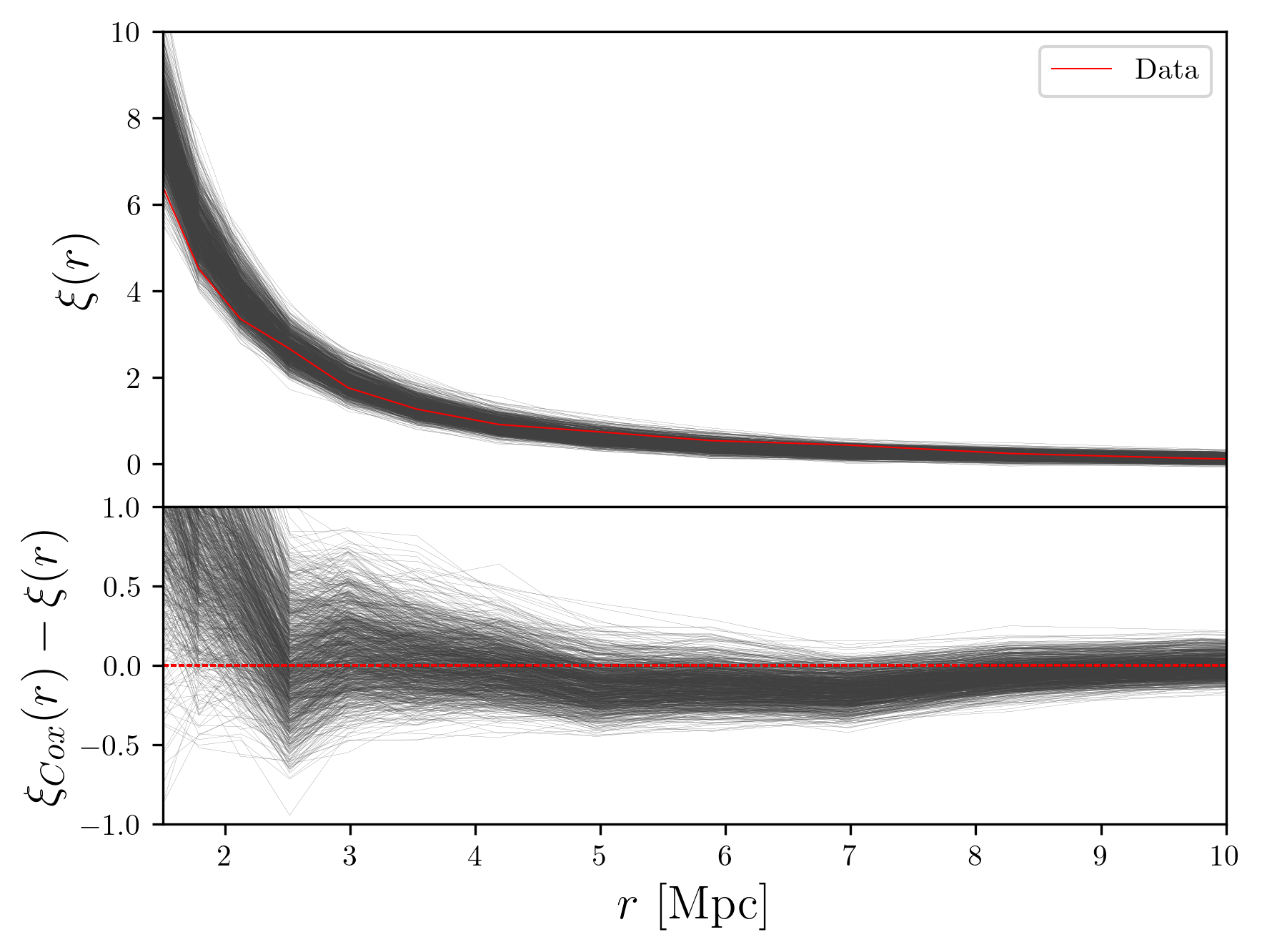}
    \end{minipage}
\caption{
\textbf{Upper panel:} 
The 2-point correlation functions for the data sample in analysis 
($\xi(r)$; red dashed line) and for the $1000$ Cox realisations ($\xi_{Cox}(r)$; black solid lines). 
\textbf{Lower panel:} 
Difference between the Cox realisations and the data 2-point correlation functions, i.e., $\xi_{Cox}(r)-\xi(r)$.
Each line corresponds to one realisation; we plotted $1000$ realisations. 
We expect the difference to be approximately equal to zero, which occurs on scales above 2 Mpc, due to the difficulties 
to adjust all scales involved in the Cox process.}
    \label{fig:diff-cox-data}
\end{figure}

\begin{figure}
    \begin{minipage}[b]{\linewidth}
    \centering
    \includegraphics[width=\textwidth]{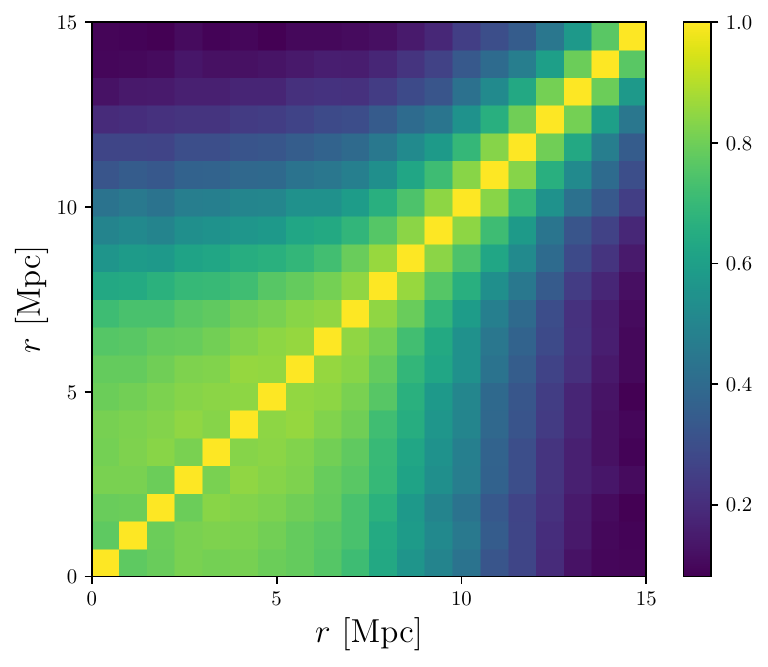}
    \end{minipage}
    \caption{Reduced covariance (or correlation) matrix, $\text{Cov}(X, Y) / (\sigma(X)\sigma(Y))$, obtained from equation~(\ref{eq:cov_matrix}) and with $\sigma(X)$, $\sigma(Y)$ being the standard deviations of $X$ and $Y$.}
    \label{fig:cov_matrix}
\end{figure}

Ultimately, the most important feature needed in these mocks catalogues is that the 2-point correlation function of the Cox realisations reproduces at the scales of interest the correlation function obtained with the HI sample data.
As a matter of fact, in figure~\ref{fig:diff-cox-data}, we show 
the difference between the Cox realisations (i.e., the mock catalogues) and the HI sample data. 
Each curve indicates the difference 
$\xi^{\text{Cox}}_{i}(r)-\xi^{\text{HI}}(r)$, where $i$ varies from 1 to 1000. As we can see, there is an excellent agreement for scales above 2 Mpc. 
This inaccuracy at small scales can be seen in the covariance matrix, 
calculated as 
\begin{equation}\label{eq:cov_matrix}
\text{Cov}(X, Y) = \frac{1}{N} \sum_{i=1}^{N} (X_i - \bar{X})(Y_i - \bar{Y}) \,,
\end{equation}
where $N$ is the sample size, and $X$ and $Y$ represent the correlation values 
at different bins (i.e., different distances between pairs). 
In figure \ref{fig:cov_matrix}, we show the reduced covariance matrix. 

In order to measure the uncertainties for the 2-point correlation function obtained using the ALFALFA data,  
we calculate the two-dimensional covariance matrix $\text{Cov}(X,Y)$ 
for the correlation function given in (\ref{2pcf}). 
Note that there is a high correlation on small scales, which decreases as we increase the scale. 
Therefore, by using the covariance matrix in our analyses, we are taking into account for the error evaluation important to quantify systematic effects of the selected sample, such as the sky surveyed area, depth of the survey, numerical density of cosmic objects, etc. 

Consequently, the covariance matrix $\text{Cov}(X,Y)$ provides a more robust measure of the relationships between observational variables, highlighting significant correlations and quantifying the influence of individual variances. 
In fact, this tool helps to identify possible correlation patterns, and is particularly useful in our study because we are interested in the analyses across different scale distances.

\section{Theoretical frame and Methodology}\label{sec:methodology}

The standard cosmological model describes the mass distribution through 
the power spectrum of matter density fluctuations, 
as a function of scale. 
This function helps to describe the evolution of matter clustering and to understand the problem of structure formation, at several scales. 
In this section, we explore fundamental concepts to measure the matter fluctuations in the ALFALFA data.

\subsection{Matter fluctuations}\label{sec:matter-flu}

The physical processes of matter clustering and structure formation 
began in the early Universe from primordial density 
fluctuations~\citep{Peebles67,FKP1994,T_Padmanabhan,Mo}. 
The evolution of these fluctuations is conveniently studied through a dimensionless quantity, the density contrast 
$\delta(\mathbf{r}, t) \equiv [\rho(\mathbf{r}, t) - \rho_{0}(t)] / \rho_0(t)$.  
Consider the matter density contrast at a given time in the early Universe, 
$\delta(\mathbf{r}) = \delta(\mathbf{r}, t_i)$, assume that at different spatial locations $\delta(\mathbf{r})$ is very weakly 
correlated, i.e., it is a random variable~\citep{T_Padmanabhan}. 
From this, one finds that its Fourier transform, $\delta_{\mathbf{k}}$, is Gaussian, therefore $\delta_{\mathbf{k}}$ is completely specified by its two moments: zero mean $\langle \delta_{\mathbf{k}} \rangle = 0$, and variance 
$\sigma_k^2 \equiv \langle |\delta_{\mathbf{k}}|^2 \rangle$, 
where $\sigma_k^2$ is the power spectrum of the density fluctuations, commonly termed $P(k)$. 
The power spectrum, $P(k)$, is the Fourier transform of the two-point correlation function, $\xi(r)$, which is a primary estimator of the large-scale matter distribution 
\begin{equation}\label{power_spectrum}
P(k) = \int \xi(r) e^{-i\mathbf{k} \cdot \mathbf{r}} \, d^3 r \,.
\end{equation}

A way to describe the clustering strength of a given ensemble of cosmic objects is to calculate the variance of the number counts within randomly placed spheres of a given radius $R$, which, considering the linear perturbations, can be written as 
\begin{equation}\label{variance}
\sigma^2(R) = \frac{1}{2\pi^2} \int P(k)\,W_{R}^{2}(k) 
\,k^2 \,dk \,,
\end{equation}
where $W_{R}$ is the top-hat window with spherical 
symmetry~\citep{Mo}. 
The variance of galaxy, or cosmic objects, distribution is used to normalize $P(k)$ in equation~(\ref{power_spectrum}). 
The function $\sigma^2(R)$ can be understood in terms of the variation in mass in a set of randomly placed windows, then one can show that the mass variance is~\citep{Mo} 
\begin{equation}\label{mass_variance}
\sigma^{2}_{M}(R) = \left\langle \left( \frac{\Delta M}{M_0} \right)^{2} \right\rangle_{\!\textbf{x}} \equiv \sigma^2(R) \,,
\end{equation}
where $\Delta M \equiv M(\textbf{x},R) - M_0(R)$, 
$M(\textbf{x},R)$ is the mass in a window centered at $\textbf{x}$, 
and $M_0(R)$ is the average of the mass defined at the scale $R$. 
It is also possible to write $\sigma^{2}(R)$ as 
\begin{equation}\label{variancia_massa}
\sigma^{2}(R)   = \int\limits_{0}^{R^{-1}} \frac{dk}{k} \Delta_{k}^2 \,,
\end{equation}
where $\Delta_{k}^2$ is a dimensionless quantity proportional to the power spectrum 
and which expresses the contribution to the variance by the power in a unit logarithmic interval of $k$~\citep{T_Padmanabhan},
\begin{equation}
\Delta_{k}^{2} \equiv \frac{k^3}{2 \pi^2} P(k) \,.
\end{equation}
It is useful to define the following quantity, which is related to $\xi(r)$ by~\citep{T_Padmanabhan} 
\begin{equation}\label{J3(r)}
J_3(R) \equiv \int_{0}^{R}r^2 \xi(r) dr \,,
\end{equation}
where this function gives us information about the spatial distribution of objects, because it represents weighted average of correlation function across different scales. Another way to represent $J_3(R)$ is 
\begin{equation}\label{J3}
J_3(R) \simeq 
\frac{R^3}{3} \int\limits_{0}^{R^{-1}} \frac{dk}{k} \Delta_{k}^2 \,,
\end{equation}
where we note that combining equations~(\ref{variancia_massa}) and (\ref{J3}) we obtain 
\begin{equation}\label{M(r)_J3}
\sigma^{2}(R) = \frac{3 J_3(R)}{R^3} \,,
\end{equation}
therefore, knowing $J_3(R)$ it is possible to calculate the mass variance at the scale $R$, i.e., $\sigma^{2}(R)$. 
Finally, one defines 
$\sigma_8 \equiv \sigma(R=8\,h^{-1}\,\text{Mpc})$, 
at the scale $R = 8\,h^{-1}$ Mpc.

\subsection{Correlation function}\label{sec:correlation-function}
We have previously seen in section~\ref{sec:matter-flu} that it is possible to relate the 2-point correlation function with the matter variance on different scales through equations~(\ref{J3(r)}) and (\ref{M(r)_J3}). 
Therefore, we shall apply the 2-point correlation function $\xi(r)$ to the ALFALFA catalogue to calculate the matter variance. 
The most widely used correlation function estimator is the Landy-Szalay~\citep{Landy}, that returns the smallest discrepancies for a given cumulative probability, has no bias and nearly Poisson variance~\citep{Martin_kerscher}. 
This estimator is given by 
\begin{equation}\label{2pcf} 
\xi(r) \equiv \frac{DD(r) - 2DR(r) + RR(r)}{RR(r)} \,,
\end{equation}
where $DD(r)$ is the number of pairs in the sample data with distance separation $r$, normalized by the total number of pairs; $RR(r)$ is a similar quantity, but for the pairs in a random sample; and $DR(r)$ corresponds to a cross-correlation between a data object and a random object. 
Using $\xi(r)$ it is possible to determine the excess probability of finding two points from a dataset at a given distance separation $r$ when compared to a random distribution. 
In practice, to obtain $\xi(r)$ we use the \textsc{TreeCorr} code~\citep{M_jarvis}~\footnote{\url{https://github.com/rmjarvis/TreeCorr}}.

\subsubsection{Power-law approximation}\label{sec:power-law}

The expectation regarding the 2-point correlation function is that it behaves according to a power-law~\citep{Peebles93,Papastergis13}, that is, 
\begin{equation}
\label{eq:power-law}
\xi(r) = \left(\frac{r}{r_0}\right)^{-\gamma} \,,
\end{equation}
where $r_0$ is a scale parameter that determines the characteristic scale of the correlation, and the parameter $\gamma$ quantifies 
the clustering of the cosmic tracer in study.

\subsubsection{Constraining Power Law Parameters Using Bayesian Inference} \label{sec:MCMC}
In order to determine the optimal power law parameters, $r_0$ and $\gamma$, from the correlation function measurements, we adopt a Bayesian analysis approach~(for more details, see e.g.,\cite{Verde10,Hobson14,Trotta17}). This involves exploring the parameter space using the Markov Chain Monte Carlo (MCMC) algorithm implemented through the publicly available 
\textsc{emcee}\footnote{\url{https://emcee.readthedocs.io/en/stable/}} code~\citep{Goodman10,Foreman13}.

In Bayesian analysis, our aim is to determine the probability of a given model with parameters $\bm{\theta}$ given a set of observations $\bm{D}$. This involves computing the posterior probability distribution $P(\bm{\theta}|\bm{D})$, which is given by 
\begin{equation}\label{eq:bayes_theorem}
P(\bm{\theta}|\bm{D}) = \frac{P(\bm{D}|\bm{\theta})P(\bm{\theta})}{P(\bm{D})}\,,
\end{equation}
where $P(\bm{D}|\bm{\theta})$ is the likelihood function, $P(\bm{\theta})$ is the prior distribution, and $P(\bm{D})$ is the evidence. As $P(\bm{D})$ can be treated like a normalisation, we can look only for the scaled posterior. The logarithm of the scaled posterior distribution can be expressed as 
\begin{equation}\label{eq:log_bayes}
\log P(\bm{\theta}|\bm{D}) \,\propto\, \log P(\bm{D}|\bm{\theta}) + \log P(\bm{\theta})\,,
\end{equation}
where the likelihood can be written as 
\begin{equation}\label{eq:log_likelihood}
\log P(\bm{D}|\bm{\theta}) \propto -\frac{1}{2}\chi^{2}\,. 
\end{equation}
In this work,
\begin{equation}\label{eq:chi_sq}
\chi^{2} = \sum_{i,j} \left[\xi(r_i) - \xi^{\text{PL}}(r_i;r_0,\gamma)\right] \times C_{i,j}^{-1}(r_i,r_j) \times \left[\xi(r_j) - \xi^{\text{PL}}(r_j;r_0,\gamma)\right] ,
\end{equation}
where $\xi(r_i)$ represents the measured correlation function, $\xi^{\text{PL}}(r_i;r_0,\gamma)$ is the model prediction with power law parameters $r_0$ and $\gamma$, and $C_{i,j}^{-1}(r_i,r_j)$ is the inverse covariance matrix of the measurements, given by equation~(\ref{eq:cov_matrix}).

In retrospect, looking at the work of \cite{Martin12}, we have properly selected the space parameter prior, as indicated in table \ref{table:priors}.

\begin{table}
\centering
\begin{tabular}{lccc}
\hline
\textbf{Parameter} & \textbf{Range} & \textbf{Initial Guess} \\
\hline
$r_0$ & $1 ~\text{Mpc} < r_0 < 15 ~\text{Mpc} $ & 5.0 \text{Mpc} \\
$\gamma$ & $1 < \gamma < 3$ & 1.8 \\
\hline
\end{tabular}
\caption{Prior ranges and initial guesses for the power-law parameters.}
\label{table:priors}
\end{table}

\section{Results and Discussion} \label{sec:results}

\begin{figure}
    \begin{minipage}[b]{\linewidth}
    \centering
    \includegraphics[width=\textwidth]{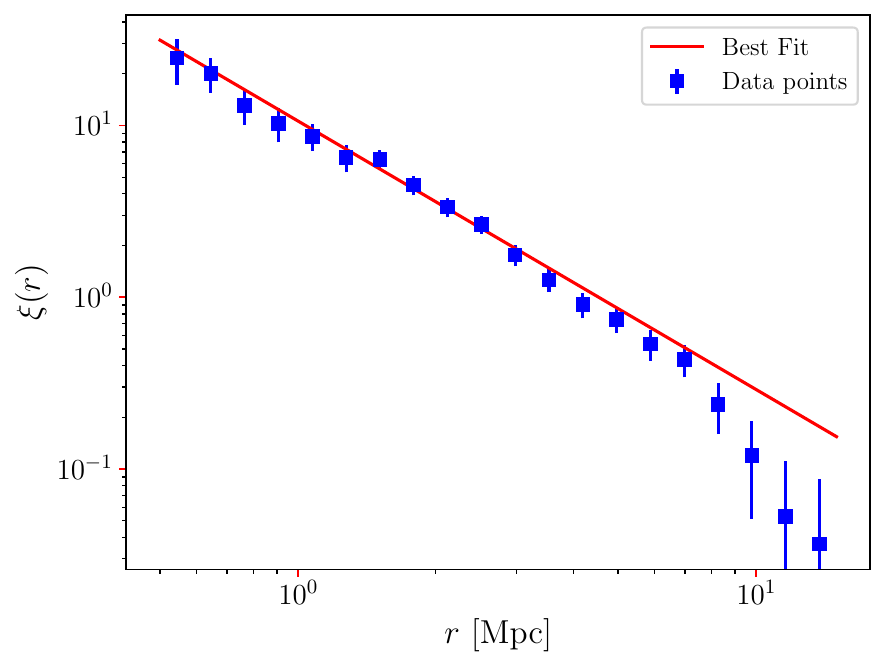}
    \end{minipage}
    \caption{2-point correlation function for the HI sample in study using, 1000 realisations of the Cox process to estimate the uncertainties. 
    For the best fit it was assumed the power-law relationship, equation~(\ref{eq:power-law}). 
    }
    \label{fig:correlation_function}
\end{figure}

Applying the statistical tools outlined in section~\ref{sec:methodology}, we analysed the selected sample described in section~\ref{sec:data}. 
Firstly, we calculated the 2-point correlation function, $\xi(r)$, for our data sample for pair distances in the interval $[0.5, 15]\,\text{Mpc}$; our result 
can be seen in figure~\ref{fig:correlation_function}, where we consider $20$ bins. 
Similarly, we applied the same procedure to a set of $1000$ mock catalogues generated from the Cox process, as detailed in section~\ref{sec:mock-catalog}, to estimate the covariance matrix using equation~(\ref{eq:cov_matrix}) and subsequently determine the uncertainties associated with our results. The findings are illustrated in figure~\ref{fig:correlation_function}.

\begin{figure}
    \begin{minipage}[b]{\linewidth}
    \centering
    \includegraphics[width=\textwidth]{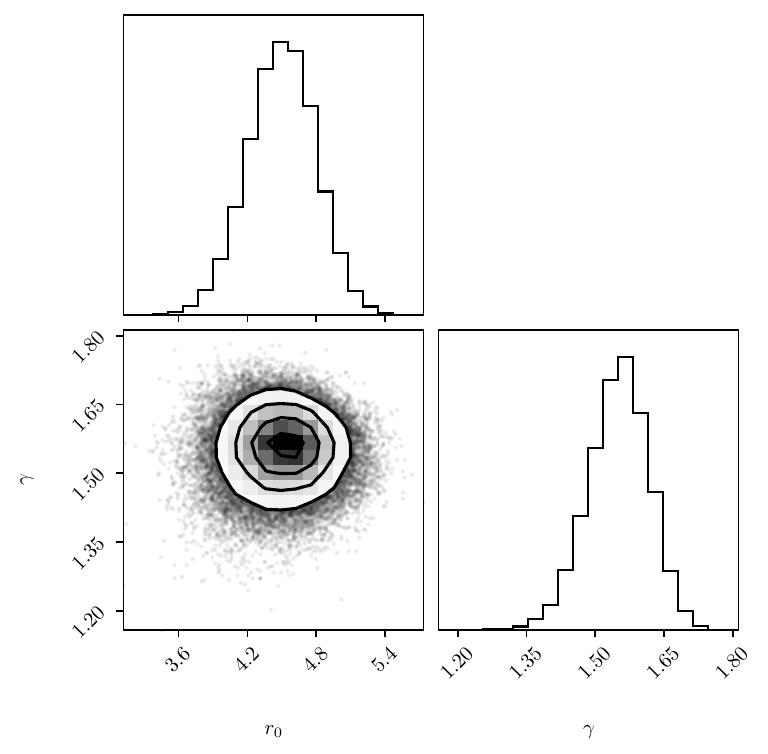}
    \end{minipage}
    \caption{Joint and marginal distributions of the power-law parameters $r_{0}$ and $\gamma$.  
    }
    \label{fig:mcmc}
\end{figure}

The best fit curve was plotted considering the power-law shown in equation~(\ref{eq:power-law}), performed over the interval $[0.5, 15]\,\text{Mpc}$, where the fitting parameters $r_{0} = 4.5 \pm 0.3\,\text{Mpc}$ and $\gamma = 1.55 \pm 0.07$ were obtained using the MCMC method (cf. section~\ref{sec:MCMC} and figure~\ref{fig:mcmc}). 
The data agree well with the expected power-law function up to 
$r \simeq 7\,\text{Mpc}$.
Beyond this scale, the observed decline in the data with respect to the best fit, is indicative of a limited sample volume, and the power-law correlation function does not suitably describe the distribution of matter structures~\citep{Martin12}. 

\begin{figure}
    \begin{minipage}[b]{\linewidth}
    \centering
    \includegraphics[width=\textwidth]{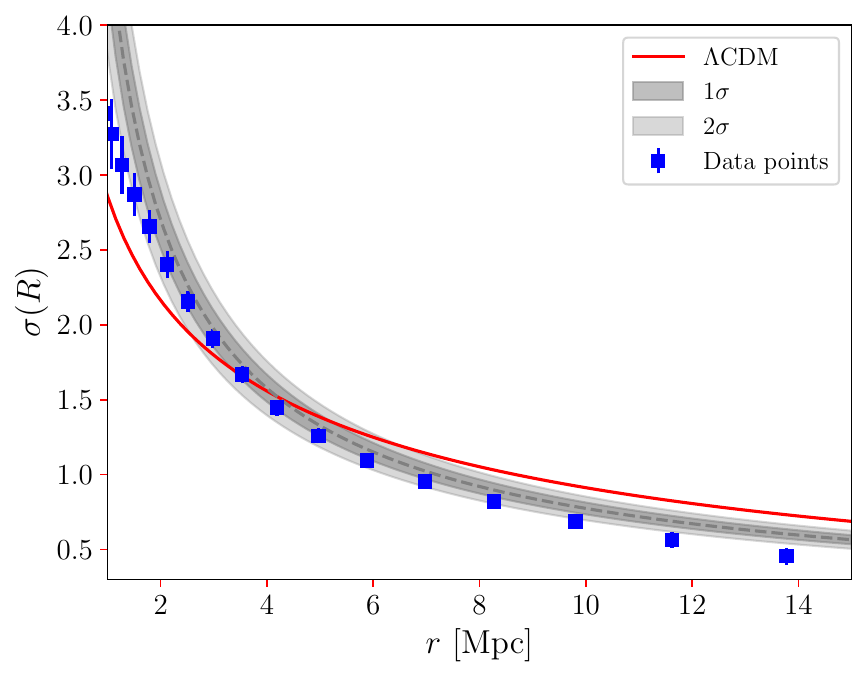}
    \end{minipage}
\caption{
Mass fluctuation for the selected HI sample from ALFALFA. The best fit was assumed to follow a power-law, equation~(\ref{eq:power-law}). 
The shaded gray areas correspond to $1\sigma$ and $2\sigma$ confidence levels. 
Note that there is an intersection between the data best-fit and the theoretical model $\Lambda$CDM, suggestive of a transition between non-linear and linear scales in the behaviour of $\sigma(R)$. 
Beyond this scale, the matter fluctuation of the HI sample consistently falls below the 
$\Lambda$CDM prediction, implying an anti-bias in our sample relative to the underlying matter.
    }
    \label{fig:mass_fluctuation}
\end{figure}

To deepen our analysis, we calculated the $\sigma(R)$ values. By using the $\xi(r)$ values and the previously obtained covariance matrix, we can get the mass fluctuation function, which is shown in figure~\ref{fig:mass_fluctuation}.
The data points were obtained directly from the square root of 
equation~(\ref{M(r)_J3}) leading to a direct measurement of the matter fluctuation of the HI extragalactic sources, 
$\sigma_{8}^{\text{tr}} = \sigma_{8}^{\text{HI}} = 0.92 \pm 0.05$ at scales $r=8$ Mpc. 
The red continuous line in figure~\ref{fig:mass_fluctuation} was generated using the Core Cosmology Library\footnote{\url{https://github.com/LSSTDESC/CCL}}~\citep[CCL;][]{Chisari2019}, which offers a robust computational tool for our calculations. 
This line corresponds to the theoretical prediction for matter fluctuations modelled under the assumption of Planck's flat-$\Lambda$CDM model~\citep{Planck20}, with the cosmological parameters specified in table~\ref{tab:pyccl}.

\begin{table}
\centering
\begin{tabular}{lccc}
\hline
\textbf{Parameter} & \textbf{Value} \\
\hline
$\Omega_c$ & $0.2656$ \\
$\Omega_b$ & $0.0494$ \\
$h$ & $0.6727$ \\
$n_s$ & $0.9649$ \\
$\sigma_8$ & $0.8120$ \\
\hline
\end{tabular}
\caption{Cosmological parameters of the Planck's flat-$\Lambda$CDM model~\citep{Planck20} used to obtain the model curve using the 
Core Cosmology Library~\citep[CCL;][]{Chisari2019}. 
See section~\ref{sec:results} for details.}
\label{tab:pyccl}
\end{table}

The dashed gray line is the best-fit expectation for the function $\sigma(R)$ when one assumes the power-law for the 2-point correlation function, because of this we call this dashed gray line as the $\sigma$-model. 
We observe a good agreement, at $2\sigma$ level, between the $\sigma$-model and the data. 
As noticed above the 2-point correlation function does not follow a power-law for scales above 7 Mpc, in consequence, this fact is translated to the 
$\sigma(R)$ computation, and manifests increasing the discrepancy between $\sigma$-model and data, as observed in figure~\ref{fig:mass_fluctuation}. 
For both the $\sigma$-model and the data, there is a transition with respect to the $\Lambda$CDM model around 4 Mpc, suggestive of a possible transition between non-linear to linear scales. 
As the scale increases, both $\sigma$-model and data fall below the $\Lambda$CDM model, consistently, indicating that the HI sources are anti-biased with respect to the underlying matter fluctuation field~\citep{Basilakos07}.

The scale where our best-fit data intersects the $\Lambda$CDM model, which results from a linear theory, is suggestive of the transition between the non-linear and linear regimes. 
We find, however, that a more appropriate scale would be to consider the scale where the matter fluctuation is $\sigma(R) = 1$, because such scale would be a model-independent quantity in data analyses where distances scales are in Mpc units 
(and not $h^{-1}$ Mpc). 
When we apply this condition to our best-fit analysis (see figure~\ref{fig:mass_fluctuation}) we obtain $R = 7.2 \pm 1.5~\text{Mpc}$. 
This could be considered as the transition scale where the linear relationship is valid $\delta_{\text{HI}}=b_{\text{HI}}\delta_{M}$, i.e., the fluctuation of matter of the HI sources is proportional to the fluctuation of underlying matter.

To determine $\sigma_8$ from our measurement, $\sigma_{8}^{\text{HI}}$, it is essential to assume a value for the linear bias, $b_{\text{HI}}$. This linear bias, in turn, requires assuming a normalisation, $\sigma_8$, for the power spectrum when performing  statistical analyses that compares models and observations~\citep{Basilakos07}. 
In summary, our $\sigma_8$ measurement serves as a consistency test for the $\Lambda$CDM model. 
However, within our approach we can study the effect of the Hubble constant on our $\sigma_8$ measurement, since one must assume a value of $h$ in order to transform  the Mpc units to $h^{-1}~\text{Mpc}$ units.

Considering the linear bias for the HI extragalactic sources obtained by \cite{Obuljen19}, $b_{\text{HI}} = 0.875 \pm 0.022$, and using the relation
\begin{equation}
\sigma_{8} = \frac{\sigma_{8}^{\text{HI}}}{b_{\text{\,HI}}}\,,
\end{equation}
with uncertainties given by
\begin{equation}
\delta \sigma_{8} = \frac{\sigma_{8}^{\text{HI}}}{b_{\text{\,HI}}} \sqrt{\left(\frac{\delta\sigma_{8}^{\text{HI}}}{\sigma_{8}^{\text{HI}}}\right)^{2} + \left(\frac{\delta b_{\text{\,HI}}}{b_{\text{\,HI}}}\right)^{2}} \,,
\end{equation}
we conclude that 
$\sigma_{8} = 1.05 \pm 0.06$ at the scale of
$8$\,Mpc. Furthermore, we consider a set of values for the Hubble parameter: 
$h = \{0.6727,\,0.698,\,0.7304\}$ to determine the $\sigma(R)$ at the 
scale $R = 8h^{-1}\,\text{Mpc}$ in each case. 
The results, with their respective uncertainties, taking into account the bias 
are listed in table~\ref{tab:sigma-bias}. 

\begin{table}
    \centering
    \begin{tabular}{lccc}
    \hline
        & $h$ & $\sigma_{8}^{\text{HI}}$ & $\sigma_{8}$\\
    \hline     
        & $0.6727$~\text{\citep{Planck20}} & $ 0.68 \pm 0.04$ & $0.78 \pm 0.04$\\
        & $0.698$~\text{\citep{Freedman21}} & $ 0.70 \pm 0.04$ & $0.80 \pm 0.05$\\
        & $0.7304~\text{\citep{Riess22}}$ & $0.73\pm 0.04$ & $0.83 \pm 0.05$\\
    \hline    
    \end{tabular}
\caption{Our measurements of 
$\sigma_{8}^{\text{HI}}$ and $\sigma_{8}$ at the scale 
$8\,h^{-1}\,\text{Mpc}$ considering diverse measurements of $h$.
}
\label{tab:sigma-bias}
\end{table}

\begin{figure}
    \begin{minipage}[b]{\linewidth}
    \centering
    \includegraphics[width=\textwidth]{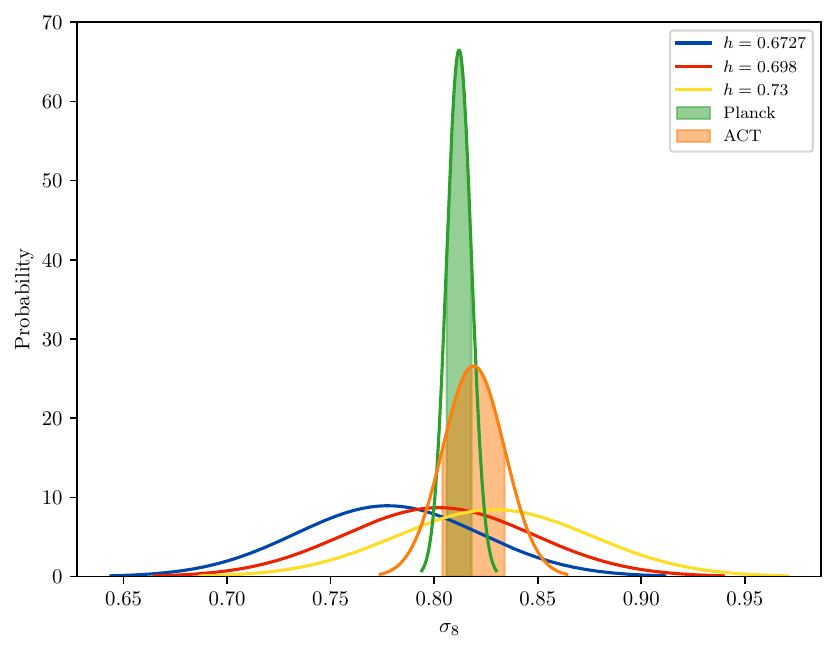}
    \end{minipage}
    \caption{Probability distribution function of $\sigma_{8}$ for different values of the Hubble parameter $h$.}
    \label{fig:prob-curves}
\end{figure}

\begin{figure}
    \begin{minipage}[b]{\linewidth}
    \centering
    \includegraphics[width=\textwidth]{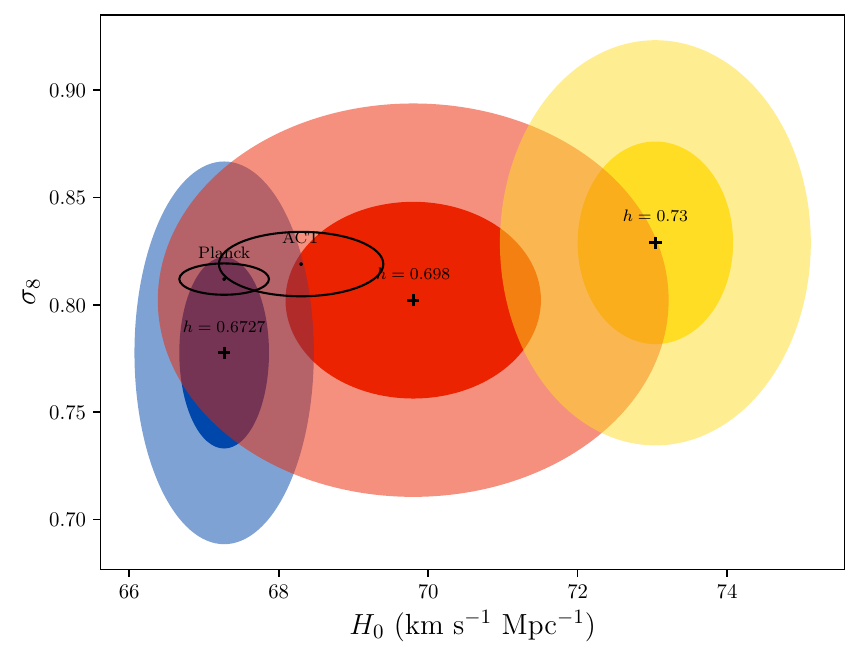}
    \end{minipage}
    \caption{$H_{0}-\sigma_{8}$ parameter space for different values of $h$. The probability ellipses show $1\sigma$ and $2\sigma$ CL. We compare our results with the ACT and Planck data.}
    \label{fig:prob-ellipses}
\end{figure}

As can be seen in the plots of the probability density functions shown in figure~\ref{fig:prob-curves} and in the $H_{0} - \sigma_{8}$ parameter space in figure~\ref{fig:prob-ellipses}, considering the tension metric between two estimates $A$ and $B$, 
\begin{equation}
T_{\sigma_{8}} = \frac{|\sigma_{8,A} - \sigma_{8,B}|}{\sqrt{\sigma_{\sigma_{8,A}}^{2} + \sigma_{\sigma_{8,B}}^{2}}} \,,
\end{equation}
we find that our measurements of $\sigma_8$ agrees at 
$0.79\,\sigma$ and $0.36\,\sigma$ confidence levels with Planck~\cite{Planck20} and Atacama Cosmology Telescope (ACT)~\cite{ACT2024} results, respectively. 

Despite $\Lambda\text{CDM}$ model best fits observational data, it still faces significant challenges, such as the tensions arising from measurements of the Hubble constant, $H_{0}$ and $\sigma_{8}$. Within the $\Lambda\text{CDM}$ framework, measurements using CMB data yield $H_{0} = 67.27 \pm 0.60$ 
km~s$^{-1}$~Mpc$^{-1}$~\citep{Planck20}, whereas local measurements using Type Ia supernovae obtain $H_{0} = 73.04 \pm 1.04$ km~s$^{-1}$~Mpc$^{-1}$~\citep{Riess22}, indicating a discrepancy of $\sim 5\sigma$ between direct and indirect measurements~\citep{Perivolaropoulos22,Valentino21b,Valentino21c,Verde23,Tully23}. Regarding $\sigma_{8}$, \cite{Planck20} reports a value of $\sigma_{8} = 0.8120 \pm 0.0073$, which is corroborated by \cite{ACT2024}, where $\sigma_{8} = 0.819 \pm 0.015$, but they both disagree with direct measurements, such as weak lensing, redshift-space distortions (RSD), and galaxy clustering at $3\sigma$ level~\citep{Perivolaropoulos22,Adhikari22,Valentino21}.

The search for alleviating these tensions motivated numerous recent studies, which report values of 
$\sigma_{8} \sim 0.75$~\citep{Freedman21,Troster20,Nunes21,Heymans21,Qu24}. 
As illustrated in figure~\ref{fig:prob-ellipses}, each value of $h$ 
provides a distinct value of $\sigma_{8}$ in units of $h^{-1}$\,Mpc. 
Our $\sigma_{8}$ measurement obtained through the ALFALFA data agrees 
at $0.79\,\sigma$ with the value reported in~\cite{Planck20}, 
using for this comparison the $h$ value provided by the Planck Collaboration. 
Similarly, when adopting the value calculated by~\cite{Freedman21}, 
$h = 0.698$, we remain $0.36\,\sigma$ away from the value reported by~\cite{ACT2024}. Some authors suggest that, despite the $\sigma_{8}$ tension potentially arising from systematic errors (in either CMB or LSS evaluation) or sample variance (originated from different regions observed by different surveys), it may indicate the necessity for a new physics beyond the standard model to explain the tension in $\sigma_{8}$ and $H_{0}$~\citep{Nunes21,Adhikari22, Perivolaropoulos22,Amon22,Schoneberg22,Poulin23,Abdalla22,Adil24,Chen24,Hollinger24,Akarsu24}.
Nevertheless, we can safely conclude that, considering the data pairs $(\sigma_8, H_0)$ from the Planck CMB and ACT CMB-lensing analyses, our $\sigma_8$ measurement agrees with them in less than $1\,\sigma$ confidence level. 

\section{Conclusions}\label{sec:conclusions}
 
The search for the cosmological model that describes the observed Universe on large scales motivate the use of astronomical data to explore alternative models through diverse 
approaches~\citep{Marques19,Edilson20,Krishnan22,Luongo22,Giare23,Giare24,Dinda24a,
Dinda24b,Bruno24,Fernanda24}. 
In this scenario, the measurement of the galaxies density fluctuation in the Universe is a crucial feature for understanding its large-scale structure, making the determination of the $\sigma_8$ parameter an important task in modern cosmology. 
It is estimated that this fluctuation in the Local Universe is approximately 1 on scales of $8 \,h^{-1}$ Mpc~\citep{Juszkiewicz10}. Over the years, various research teams have come together to measure the matter fluctuations using diverse techniques and cosmic tracers~\citep{Hasselfield,Reid,Melchiorri,Hamana}, seeking a better understanding of the tendency for galaxies to cluster more densely than the underlying dark matter and the amplitude of primordial fluctuations in the Universe, resulting from physical processes during its early stages. The amplitude of mass fluctuations describes the normalization of the linear spectrum of mass fluctuations in the early Universe, as described in section~\ref{sec:matter-flu}, the spectrum that seeded galaxy formation. 
Thus, massive clusters form from outlier fluctuations of the primordial Gaussian density field, 
this means that their abundance is exponentially sensitive to the overall 
normalisation~\citep{Bahcall}.

The increasing tension surrounding the value of $\sigma_8$ necessitates additional studies and analyses with varied datasets and different methodologies to obtain a more precise and comprehensive understanding of matter fluctuation amplitudes in the Universe. In our work, we analyzed the large-scale structure of the Universe using data from the ALFALFA extragalactic HI survey. Our focus was on characterizing matter fluctuations and understanding mass variance on different spatial scales, utilizing the HI extragalactic sources as tracers of the underlying matter distribution.

We investigated the matter fluctuation $\sigma(R)$ by applying the 2-point correlation function to a selected sample of the ALFALFA survey, and using the relation between this quantity and $J_{3}(R)$. 
The derived value of $\sigma_{8}^{\text{HI}} = 0.92 \pm 0.05$ aligns with the expectations for galaxy distribution and is consistent with the prediction of the $\Lambda\text{CDM}$ model for a sample with bias factor $b < 1$. When we consider the bias for HI sources, $b_{\text{HI}}$, this yields $\sigma_{8} = 1.05 \pm 0.06$. Considering different values for the Hubble parameter, we found that each value of $h$ provides a distinct value of $\sigma_{8}$, which is different from the previous measurements from Planck and ACT at $0.79\sigma$ and $0.36\sigma$, respectively.

Our results obtained  from the analyses of the ALFALFA data highlight the robustness of our methodology to measure the matter fluctuations in the Local Universe, potentially contributing to the ongoing discussion concerning the $\sigma_{8}$, or $S_8$, tension. 

\section*{Acknowledgements}
CF and ML thank Coordenação de Aperfeiçoamento de Pessoal de Nível Superior (CAPES),
JO and AB acknowledge to Conselho Nacional de Desenvolvimento Científico e Tecnológico (CNPq), for their corresponding fellowships. 
FA thanks CNPq and Fundação Carlos Chagas Filho de Amparo à Pesquisa do Estado do Rio de Janeiro (FAPERJ), Processo SEI 260003/014913/2023, for the financial support.

\section*{Data Availability}
The data underlying this article will be shared on reasonable request to the corresponding author.



\bibliographystyle{mnras}
\bibliography{bib} 

\begin{thebibliography}{}
\makeatletter
\relax
\def\mn@urlcharsother{\let\do\@makeother \do\$\do\&\do\#\do\^\do\_\do\%\do\~}
\def\mn@doi{\begingroup\mn@urlcharsother \@ifnextchar [ {\mn@doi@} {\mn@doi@[]}}
\def\mn@doi@[#1]#2{\def\@tempa{#1}\ifx\@tempa\@empty \href {http://dx.doi.org/#2} {doi:#2}\else \href {http://dx.doi.org/#2} {#1}\fi \endgroup}
\def\mn@eprint#1#2{\mn@eprint@#1:#2::\@nil}
\def\mn@eprint@arXiv#1{\href {http://arxiv.org/abs/#1} {{\tt arXiv:#1}}}
\def\mn@eprint@dblp#1{\href {http://dblp.uni-trier.de/rec/bibtex/#1.xml} {dblp:#1}}
\def\mn@eprint@#1:#2:#3:#4\@nil{\def\@tempa {#1}\def\@tempb {#2}\def\@tempc {#3}\ifx \@tempc \@empty \let \@tempc \@tempb \let \@tempb \@tempa \fi \ifx \@tempb \@empty \def\@tempb {arXiv}\fi \@ifundefined {mn@eprint@\@tempb}{\@tempb:\@tempc}{\expandafter \expandafter \csname mn@eprint@\@tempb\endcsname \expandafter{\@tempc}}}

\bibitem[\protect\citeauthoryear{{Abdalla} et~al.,}{{Abdalla} et~al.}{2022}]{Abdalla22}
{Abdalla} E.,  et~al., 2022, \mn@doi [Journal of High Energy Astrophysics] {10.1016/j.jheap.2022.04.002}, \href {https://ui.adsabs.harvard.edu/abs/2022JHEAp..34...49A} {34, 49}

\bibitem[\protect\citeauthoryear{{Adhikari} et~al.,}{{Adhikari} et~al.}{2022}]{Adhikari22}
{Adhikari} R.~X.,  et~al., 2022, \mn@doi [arXiv e-prints] {10.48550/arXiv.2209.11726}, \href {https://ui.adsabs.harvard.edu/abs/2022arXiv220911726A} {p. arXiv:2209.11726}

\bibitem[\protect\citeauthoryear{{Adil}, {Akarsu}, {Malekjani}, {{\'O} Colg{\'a}in}, {Pourojaghi}, {Sen}  \& {Sheikh-Jabbari}}{{Adil} et~al.}{2024}]{Adil24}
{Adil} S.~A.,  {Akarsu} {\"O}.,  {Malekjani} M.,  {{\'O} Colg{\'a}in} E.,  {Pourojaghi} S.,  {Sen} A.~A.,   {Sheikh-Jabbari} M.~M.,  2024, \mn@doi [\mnras] {10.1093/mnrasl/slad165}, \href {https://ui.adsabs.harvard.edu/abs/2024MNRAS.528L..20A} {528, L20}

\bibitem[\protect\citeauthoryear{{Akarsu}, {Colg{\'a}in}, {Sen}  \& {Sheikh-Jabbari}}{{Akarsu} et~al.}{2024}]{Akarsu24}
{Akarsu} {\"O}.,  {Colg{\'a}in} E.~{\'O}.,  {Sen} A.~A.,   {Sheikh-Jabbari} M.~M.,  2024, \mn@doi [arXiv e-prints] {10.48550/arXiv.2410.23134}, \href {https://ui.adsabs.harvard.edu/abs/2024arXiv241023134A} {p. arXiv:2410.23134}

\bibitem[\protect\citeauthoryear{{Amon} \& {Efstathiou}}{{Amon} \& {Efstathiou}}{2022}]{Amon22}
{Amon} A.,  {Efstathiou} G.,  2022, \mn@doi [\mnras] {10.1093/mnras/stac2429}, \href {https://ui.adsabs.harvard.edu/abs/2022MNRAS.516.5355A} {516, 5355}

\bibitem[\protect\citeauthoryear{{Avila}, {Novaes}, {Bernui}  \& {de Carvalho}}{{Avila} et~al.}{2018}]{Avila18}
{Avila} F.,  {Novaes} C.~P.,  {Bernui} A.,   {de Carvalho} E.,  2018, \mn@doi [\jcap] {10.1088/1475-7516/2018/12/041}, \href {https://ui.adsabs.harvard.edu/abs/2018JCAP...12..041A} {2018, 041}

\bibitem[\protect\citeauthoryear{{Avila}, {Novaes}, {Bernui}, {de Carvalho}  \& {Nogueira-Cavalcante}}{{Avila} et~al.}{2019}]{Avila19}
{Avila} F.,  {Novaes} C.~P.,  {Bernui} A.,  {de Carvalho} E.,   {Nogueira-Cavalcante} J.~P.,  2019, \mn@doi [\mnras] {10.1093/mnras/stz1765}, \href {https://ui.adsabs.harvard.edu/abs/2019MNRAS.488.1481A} {488, 1481}

\bibitem[\protect\citeauthoryear{{Avila}, {Bernui}, {de Carvalho}  \& {Novaes}}{{Avila} et~al.}{2021}]{Avila21}
{Avila} F.,  {Bernui} A.,  {de Carvalho} E.,   {Novaes} C.~P.,  2021, \mn@doi [\mnras] {10.1093/mnras/stab1488}, \href {https://ui.adsabs.harvard.edu/abs/2021MNRAS.505.3404A} {505, 3404}

\bibitem[\protect\citeauthoryear{{Avila}, {Bernui}, {Bonilla}  \& {Nunes}}{{Avila} et~al.}{2022}]{Avila2022}
{Avila} F.,  {Bernui} A.,  {Bonilla} A.,   {Nunes} R.~C.,  2022, \mn@doi [European Physical Journal C] {10.1140/epjc/s10052-022-10561-0}, \href {https://ui.adsabs.harvard.edu/abs/2022EPJC...82..594A} {82, 594}

\bibitem[\protect\citeauthoryear{{Avila}, {Oliveira}, {Dias}  \& {Bernui}}{{Avila} et~al.}{2023}]{Avila23}
{Avila} F.,  {Oliveira} J.,  {Dias} M. L.~S.,   {Bernui} A.,  2023, \mn@doi [Brazilian Journal of Physics] {10.1007/s13538-023-01259-z}, \href {https://ui.adsabs.harvard.edu/abs/2023BrJPh..53...49A} {53, 49}

\bibitem[\protect\citeauthoryear{{Avila}, {de Carvalho}, {Bernui}, {Lima}  \& {Nunes}}{{Avila} et~al.}{2024}]{Avila24}
{Avila} F.,  {de Carvalho} E.,  {Bernui} A.,  {Lima} H.,   {Nunes} R.~C.,  2024, \mn@doi [\mnras] {10.1093/mnras/stae867}, \href {https://ui.adsabs.harvard.edu/abs/2024MNRAS.529.4980A} {529, 4980}

\bibitem[\protect\citeauthoryear{{Bahcall} \& {Bode}}{{Bahcall} \& {Bode}}{2003}]{Bahcall}
{Bahcall} N.~A.,  {Bode} P.,  2003, \mn@doi [\apjl] {10.1086/375503}, \href {https://ui.adsabs.harvard.edu/abs/2003ApJ...588L...1B} {588, L1}

\bibitem[\protect\citeauthoryear{{Basilakos}, {Plionis}, {Kova{\v{c}}}  \& {Voglis}}{{Basilakos} et~al.}{2007}]{Basilakos07}
{Basilakos} S.,  {Plionis} M.,  {Kova{\v{c}}} K.,   {Voglis} N.,  2007, \mn@doi [\mnras] {10.1111/j.1365-2966.2007.11781.x}, \href {https://ui.adsabs.harvard.edu/abs/2007MNRAS.378..301B} {378, 301}

\bibitem[\protect\citeauthoryear{{Borgani}, {Plionis}, {Coles}  \& {Moscardini}}{{Borgani} et~al.}{1995}]{Borgani95}
{Borgani} S.,  {Plionis} M.,  {Coles} P.,   {Moscardini} L.,  1995, \mn@doi [\mnras] {10.1093/mnras/277.4.1191}, \href {https://ui.adsabs.harvard.edu/abs/1995MNRAS.277.1191B} {277, 1191}

\bibitem[\protect\citeauthoryear{Chalela}{Chalela}{2021}]{Chalela_RandomSDSS_2021}
Chalela M.,  2021, {RandomSDSS}, \url {https://github.com/mchalela/RandomSDSS}

\bibitem[\protect\citeauthoryear{{Chen}, {Ivanov}, {Philcox}  \& {Wenzl}}{{Chen} et~al.}{2024}]{Chen24}
{Chen} S.-F.,  {Ivanov} M.~M.,  {Philcox} O. H.~E.,   {Wenzl} L.,  2024, \mn@doi [\prl] {10.1103/PhysRevLett.133.231001}, \href {https://ui.adsabs.harvard.edu/abs/2024PhRvL.133w1001C} {133, 231001}

\bibitem[\protect\citeauthoryear{{Chisari} et~al.,}{{Chisari} et~al.}{2019}]{Chisari2019}
{Chisari} N.~E.,  et~al., 2019, \mn@doi [\apjs] {10.3847/1538-4365/ab1658}, \href {https://ui.adsabs.harvard.edu/abs/2019ApJS..242....2C} {242, 2}

\bibitem[\protect\citeauthoryear{{Di Valentino} et~al.,}{{Di Valentino} et~al.}{2021a}]{Valentino21b}
{Di Valentino} E.,  et~al., 2021a, \mn@doi [Classical and Quantum Gravity] {10.1088/1361-6382/ac086d}, \href {https://ui.adsabs.harvard.edu/abs/2021CQGra..38o3001D} {38, 153001}

\bibitem[\protect\citeauthoryear{{Di Valentino} et~al.,}{{Di Valentino} et~al.}{2021b}]{Valentino21}
{Di Valentino} E.,  et~al., 2021b, \mn@doi [Astroparticle Physics] {10.1016/j.astropartphys.2021.102604}, \href {https://ui.adsabs.harvard.edu/abs/2021APh...13102604D} {131, 102604}

\bibitem[\protect\citeauthoryear{{Di Valentino} et~al.,}{{Di Valentino} et~al.}{2021c}]{Valentino21c}
{Di Valentino} E.,  et~al., 2021c, \mn@doi [Astroparticle Physics] {10.1016/j.astropartphys.2021.102605}, \href {https://ui.adsabs.harvard.edu/abs/2021APh...13102605D} {131, 102605}

\bibitem[\protect\citeauthoryear{{Dinda}}{{Dinda}}{2024}]{Dinda24b}
{Dinda} B.~R.,  2024, \mn@doi [European Physical Journal C] {10.1140/epjc/s10052-024-12774-x}, \href {https://ui.adsabs.harvard.edu/abs/2024EPJC...84..402D} {84, 402}

\bibitem[\protect\citeauthoryear{{Dinda} \& {Banerjee}}{{Dinda} \& {Banerjee}}{2024}]{Dinda24a}
{Dinda} B.~R.,  {Banerjee} N.,  2024, \mn@doi [European Physical Journal C] {10.1140/epjc/s10052-024-13064-2}, \href {https://ui.adsabs.harvard.edu/abs/2024EPJC...84..688D} {84, 688}

\bibitem[\protect\citeauthoryear{{Efstathiou}}{{Efstathiou}}{1995}]{Efstathiou95}
{Efstathiou} G.,  1995, \mn@doi [\mnras] {10.1093/mnras/276.4.1425}, \href {https://ui.adsabs.harvard.edu/abs/1995MNRAS.276.1425E} {276, 1425}

\bibitem[\protect\citeauthoryear{{Efstathiou}, {Kaiser}, {Saunders}, {Lawrence}, {Rowan-Robinson}, {Ellis}  \& {Frenk}}{{Efstathiou} et~al.}{1990}]{Efstathiou90}
{Efstathiou} G.,  {Kaiser} N.,  {Saunders} W.,  {Lawrence} A.,  {Rowan-Robinson} M.,  {Ellis} R.~S.,   {Frenk} C.~S.,  1990, \mnras, \href {https://ui.adsabs.harvard.edu/abs/1990MNRAS.247P..10E} {247, 10P}

\bibitem[\protect\citeauthoryear{{Feldman}, {Kaiser}  \& {Peacock}}{{Feldman} et~al.}{1994}]{FKP1994}
{Feldman} H.~A.,  {Kaiser} N.,   {Peacock} J.~A.,  1994, \mn@doi [\apj] {10.1086/174036}, \href {https://ui.adsabs.harvard.edu/abs/1994ApJ...426...23F} {426, 23}

\bibitem[\protect\citeauthoryear{{Foreman-Mackey}, {Hogg}, {Lang}  \& {Goodman}}{{Foreman-Mackey} et~al.}{2013}]{Foreman13}
{Foreman-Mackey} D.,  {Hogg} D.~W.,  {Lang} D.,   {Goodman} J.,  2013, \mn@doi [\pasp] {10.1086/670067}, \href {https://ui.adsabs.harvard.edu/abs/2013PASP..125..306F} {125, 306}

\bibitem[\protect\citeauthoryear{{Franco}, {Avila}  \& {Bernui}}{{Franco} et~al.}{2024}]{Franco24}
{Franco} C.,  {Avila} F.,   {Bernui} A.,  2024, \mn@doi [\mnras] {10.1093/mnras/stad3616}, \href {https://ui.adsabs.harvard.edu/abs/2024MNRAS.527.7400F} {527, 7400}

\bibitem[\protect\citeauthoryear{{Freedman}}{{Freedman}}{2021}]{Freedman21}
{Freedman} W.~L.,  2021, \mn@doi [\apj] {10.3847/1538-4357/ac0e95}, \href {https://ui.adsabs.harvard.edu/abs/2021ApJ...919...16F} {919, 16}

\bibitem[\protect\citeauthoryear{{Gausmann} \& {Ferrari}}{{Gausmann} \& {Ferrari}}{2024}]{Gausmann2024}
{Gausmann} E.,  {Ferrari} F.,  2024, \mn@doi [arXiv e-prints] {10.48550/arXiv.2407.02213}, \href {https://ui.adsabs.harvard.edu/abs/2024arXiv240702213G} {p. arXiv:2407.02213}

\bibitem[\protect\citeauthoryear{{Giar{\`e}}}{{Giar{\`e}}}{2024}]{Giare24}
{Giar{\`e}} W.,  2024, \mn@doi [\prd] {10.1103/PhysRevD.109.123545}, \href {https://ui.adsabs.harvard.edu/abs/2024PhRvD.109l3545G} {109, 123545}

\bibitem[\protect\citeauthoryear{{Giar{\`e}}, {G{\'o}mez-Valent}, {Di Valentino}  \& {van de Bruck}}{{Giar{\`e}} et~al.}{2024}]{Giare23}
{Giar{\`e}} W.,  {G{\'o}mez-Valent} A.,  {Di Valentino} E.,   {van de Bruck} C.,  2024, \mn@doi [\prd] {10.1103/PhysRevD.109.063516}, \href {https://ui.adsabs.harvard.edu/abs/2024PhRvD.109f3516G} {109, 063516}

\bibitem[\protect\citeauthoryear{{Goodman} \& {Weare}}{{Goodman} \& {Weare}}{2010}]{Goodman10}
{Goodman} J.,  {Weare} J.,  2010, \mn@doi [Communications in Applied Mathematics and Computational Science] {10.2140/camcos.2010.5.65}, \href {https://ui.adsabs.harvard.edu/abs/2010CAMCS...5...65G} {5, 65}

\bibitem[\protect\citeauthoryear{{Hamana} et~al.,}{{Hamana} et~al.}{2003}]{Hamana}
{Hamana} T.,  et~al., 2003, \mn@doi [\apj] {10.1086/378348}, \href {https://ui.adsabs.harvard.edu/abs/2003ApJ...597...98H} {597, 98}

\bibitem[\protect\citeauthoryear{{Hasselfield}, {Hilton}, {Marriage}, {Addison}  \& {Barrientos}}{{Hasselfield} et~al.}{2013}]{Hasselfield}
{Hasselfield} M.,  {Hilton} M.,  {Marriage} T.~A.,  {Addison} G.~E.,   {Barrientos} E.~J.,  2013, \mn@doi [\jcap] {10.1088/1475-7516/2013/07/008}, \href {https://ui.adsabs.harvard.edu/abs/2013JCAP...07..008H} {2013, 008}

\bibitem[\protect\citeauthoryear{{Haynes} et~al.,}{{Haynes} et~al.}{2018}]{Haynes18}
{Haynes} M.~P.,  et~al., 2018, \mn@doi [\apj] {10.3847/1538-4357/aac956}, \href {https://ui.adsabs.harvard.edu/abs/2018ApJ...861...49H} {861, 49}

\bibitem[\protect\citeauthoryear{{Heymans} et~al.,}{{Heymans} et~al.}{2021}]{Heymans21}
{Heymans} C.,  et~al., 2021, \mn@doi [\aap] {10.1051/0004-6361/202039063}, \href {https://ui.adsabs.harvard.edu/abs/2021A&A...646A.140H} {646, A140}

\bibitem[\protect\citeauthoryear{{Hobson}, {Jaffe}, {Liddle}, {Mukherjee}  \& {Parkinson}}{{Hobson} et~al.}{2014}]{Hobson14}
{Hobson} M.~P.,  {Jaffe} A.~H.,  {Liddle} A.~R.,  {Mukherjee} P.,   {Parkinson} D.,  2014, {Bayesian Methods in Cosmology}.
Cambridge University Press

\bibitem[\protect\citeauthoryear{{Hollinger} \& {Hudson}}{{Hollinger} \& {Hudson}}{2024}]{Hollinger24}
{Hollinger} A.~M.,  {Hudson} M.~J.,  2024, \mn@doi [\mnras] {10.1093/mnras/stae1042}, \href {https://ui.adsabs.harvard.edu/abs/2024MNRAS.531..788H} {531, 788}

\bibitem[\protect\citeauthoryear{{Jarvis}, {Bernstein}  \& {Jain}}{{Jarvis} et~al.}{2004}]{M_jarvis}
{Jarvis} M.,  {Bernstein} G.,   {Jain} B.,  2004, \mn@doi [\mnras] {10.1111/j.1365-2966.2004.07926.x}, \href {https://ui.adsabs.harvard.edu/abs/2004MNRAS.352..338J} {352, 338}

\bibitem[\protect\citeauthoryear{{Jones}, {Papastergis}, {Haynes}  \& {Giovanelli}}{{Jones} et~al.}{2016}]{Jones16}
{Jones} M.~G.,  {Papastergis} E.,  {Haynes} M.~P.,   {Giovanelli} R.,  2016, \mn@doi [\mnras] {10.1093/mnras/stw263}, \href {https://ui.adsabs.harvard.edu/abs/2016MNRAS.457.4393J} {457, 4393}

\bibitem[\protect\citeauthoryear{{Jones}, {Haynes}, {Giovanelli}  \& {Moorman}}{{Jones} et~al.}{2018}]{Jones18}
{Jones} M.~G.,  {Haynes} M.~P.,  {Giovanelli} R.,   {Moorman} C.,  2018, \mn@doi [\mnras] {10.1093/mnras/sty521}, \href {https://ui.adsabs.harvard.edu/abs/2018MNRAS.477....2J} {477, 2}

\bibitem[\protect\citeauthoryear{{Juszkiewicz}, {Feldman}, {Fry}  \& {Jaffe}}{{Juszkiewicz} et~al.}{2010}]{Juszkiewicz10}
{Juszkiewicz} R.,  {Feldman} H.~A.,  {Fry} J.~N.,   {Jaffe} A.~H.,  2010, \mn@doi [\jcap] {10.1088/1475-7516/2010/02/021}, \href {https://ui.adsabs.harvard.edu/abs/2010JCAP...02..021J} {2010, 021}

\bibitem[\protect\citeauthoryear{{Kaiser}}{{Kaiser}}{1988}]{kaiser88}
{Kaiser} N.,  1988, \mn@doi [\mnras] {10.1093/mnras/231.2.149}, \href {https://ui.adsabs.harvard.edu/abs/1988MNRAS.231..149K} {231, 149}

\bibitem[\protect\citeauthoryear{{Kalbouneh}, {Marinoni}  \& {Bel}}{{Kalbouneh} et~al.}{2023}]{Marinoni2023}
{Kalbouneh} B.,  {Marinoni} C.,   {Bel} J.,  2023, \mn@doi [\prd] {10.1103/PhysRevD.107.023507}, \href {https://ui.adsabs.harvard.edu/abs/2023PhRvD.107b3507K} {107, 023507}

\bibitem[\protect\citeauthoryear{{Keih{\"a}nen} et~al.,}{{Keih{\"a}nen} et~al.}{2019}]{Keihanen19}
{Keih{\"a}nen} E.,  et~al., 2019, \mn@doi [\aap] {10.1051/0004-6361/201935828}, \href {https://ui.adsabs.harvard.edu/abs/2019A&A...631A..73K} {631, A73}

\bibitem[\protect\citeauthoryear{{Kerscher}, {Szapudi}  \& {Szalay}}{{Kerscher} et~al.}{2000}]{Martin_kerscher}
{Kerscher} M.,  {Szapudi} I.,   {Szalay} A.~S.,  2000, \mn@doi [\apjl] {10.1086/312702}, \href {https://ui.adsabs.harvard.edu/abs/2000ApJ...535L..13K} {535, L13}

\bibitem[\protect\citeauthoryear{{Krishnan}, {Mohayaee}, {Colg{\'a}in}, {Sheikh-Jabbari}  \& {Yin}}{{Krishnan} et~al.}{2022}]{Krishnan22}
{Krishnan} C.,  {Mohayaee} R.,  {Colg{\'a}in} E.~{\'O}.,  {Sheikh-Jabbari} M.~M.,   {Yin} L.,  2022, \mn@doi [\prd] {10.1103/PhysRevD.105.063514}, \href {https://ui.adsabs.harvard.edu/abs/2022PhRvD.105f3514K} {105, 063514}

\bibitem[\protect\citeauthoryear{{Landy} \& {Szalay}}{{Landy} \& {Szalay}}{1993}]{Landy}
{Landy} S.~D.,  {Szalay} A.~S.,  1993, \mn@doi [\apj] {10.1086/172900}, \href {https://ui.adsabs.harvard.edu/abs/1993ApJ...412...64L} {412, 64}

\bibitem[\protect\citeauthoryear{{Lopes}, {Bernui}, {Franco}  \& {Avila}}{{Lopes} et~al.}{2024}]{Lopes24}
{Lopes} M.,  {Bernui} A.,  {Franco} C.,   {Avila} F.,  2024, \mn@doi [\apj] {10.3847/1538-4357/ad3735}, \href {https://ui.adsabs.harvard.edu/abs/2024ApJ...967...47L} {967, 47}

\bibitem[\protect\citeauthoryear{{Luongo}, {Muccino}, {Colg{\'a}in}, {Sheikh-Jabbari}  \& {Yin}}{{Luongo} et~al.}{2022}]{Luongo22}
{Luongo} O.,  {Muccino} M.,  {Colg{\'a}in} E.~{\'O}.,  {Sheikh-Jabbari} M.~M.,   {Yin} L.,  2022, \mn@doi [\prd] {10.1103/PhysRevD.105.103510}, \href {https://ui.adsabs.harvard.edu/abs/2022PhRvD.105j3510L} {105, 103510}

\bibitem[\protect\citeauthoryear{{Madhavacheril} et~al.,}{{Madhavacheril} et~al.}{2024}]{ACT2024}
{Madhavacheril} M.~S.,  et~al., 2024, \mn@doi [\apj] {10.3847/1538-4357/acff5f}, \href {https://ui.adsabs.harvard.edu/abs/2024ApJ...962..113M} {962, 113}

\bibitem[\protect\citeauthoryear{{Marques}, {Liu}, {Zorrilla Matilla}, {Haiman}, {Bernui}  \& {Novaes}}{{Marques} et~al.}{2019}]{Marques19}
{Marques} G.~A.,  {Liu} J.,  {Zorrilla Matilla} J.~M.,  {Haiman} Z.,  {Bernui} A.,   {Novaes} C.~P.,  2019, \mn@doi [\jcap] {10.1088/1475-7516/2019/06/019}, \href {https://ui.adsabs.harvard.edu/abs/2019JCAP...06..019M} {06, 019}

\bibitem[\protect\citeauthoryear{{Martin}, {Giovanelli}, {Haynes}  \& {Guzzo}}{{Martin} et~al.}{2012}]{Martin12}
{Martin} A.~M.,  {Giovanelli} R.,  {Haynes} M.~P.,   {Guzzo} L.,  2012, \mn@doi [\apj] {10.1088/0004-637X/750/1/38}, \href {https://ui.adsabs.harvard.edu/abs/2012ApJ...750...38M} {750, 38}

\bibitem[\protect\citeauthoryear{Martinez \& Saar}{Martinez \& Saar}{2001}]{Martinez01}
Martinez V.~J.,  Saar E.,  2001, Statistics of the galaxy distribution.
Chapman and Hall/CRC, Boca Raton, Florida

\bibitem[\protect\citeauthoryear{{Martinez}, {Pons-Borderia}, {Moyeed}  \& {Graham}}{{Martinez} et~al.}{1998}]{Martinez98}
{Martinez} V.~J.,  {Pons-Borderia} M.-J.,  {Moyeed} R.~A.,   {Graham} M.~J.,  1998, \mn@doi [\mnras] {10.1046/j.1365-8711.1998.01730.x}, \href {https://ui.adsabs.harvard.edu/abs/1998MNRAS.298.1212M} {298, 1212}

\bibitem[\protect\citeauthoryear{{Masters}}{{Masters}}{2005}]{Masters2005}
{Masters} K.~L.,  2005, PhD thesis, Cornell University, New York

\bibitem[\protect\citeauthoryear{{Melchiorri}, {Bode}, {Bahcall}  \& {Silk}}{{Melchiorri} et~al.}{2003}]{Melchiorri}
{Melchiorri} A.,  {Bode} P.,  {Bahcall} N.~A.,   {Silk} J.,  2003, \mn@doi [\apjl] {10.1086/374584}, \href {https://ui.adsabs.harvard.edu/abs/2003ApJ...586L...1M} {586, L1}

\bibitem[\protect\citeauthoryear{{Migkas}, {Pacaud}, {Schellenberger}, {Erler}, {Nguyen-Dang}, {Reiprich}, {Ramos-Ceja}  \& {Lovisari}}{{Migkas} et~al.}{2021}]{Migkas2021}
{Migkas} K.,  {Pacaud} F.,  {Schellenberger} G.,  {Erler} J.,  {Nguyen-Dang} N.~T.,  {Reiprich} T.~H.,  {Ramos-Ceja} M.~E.,   {Lovisari} L.,  2021, \mn@doi [\aap] {10.1051/0004-6361/202140296}, \href {https://ui.adsabs.harvard.edu/abs/2021A&A...649A.151M} {649, A151}

\bibitem[\protect\citeauthoryear{Mo, Van~den Bosch  \& White}{Mo et~al.}{2010}]{Mo}
Mo H.,  Van~den Bosch F.,   White S.,  2010, Galaxy formation and evolution.
Cambridge University Press

\bibitem[\protect\citeauthoryear{{Norberg}, {Baugh}, {Gazta{\~n}aga}  \& {Croton}}{{Norberg} et~al.}{2009}]{Norberg09}
{Norberg} P.,  {Baugh} C.~M.,  {Gazta{\~n}aga} E.,   {Croton} D.~J.,  2009, \mn@doi [\mnras] {10.1111/j.1365-2966.2009.14389.x}, \href {https://ui.adsabs.harvard.edu/abs/2009MNRAS.396...19N} {396, 19}

\bibitem[\protect\citeauthoryear{{Nunes} \& {Vagnozzi}}{{Nunes} \& {Vagnozzi}}{2021}]{Nunes21}
{Nunes} R.~C.,  {Vagnozzi} S.,  2021, \mn@doi [\mnras] {10.1093/mnras/stab1613}, \href {https://ui.adsabs.harvard.edu/abs/2021MNRAS.505.5427N} {505, 5427}

\bibitem[\protect\citeauthoryear{{Obuljen}, {Alonso}, {Villaescusa-Navarro}, {Yoon}  \& {Jones}}{{Obuljen} et~al.}{2019}]{Obuljen19}
{Obuljen} A.,  {Alonso} D.,  {Villaescusa-Navarro} F.,  {Yoon} I.,   {Jones} M.,  2019, \mn@doi [\mnras] {10.1093/mnras/stz1118}, \href {https://ui.adsabs.harvard.edu/abs/2019MNRAS.486.5124O} {486, 5124}

\bibitem[\protect\citeauthoryear{{Oliveira}, {Avila}, {Bernui}, {Bonilla}  \& {Nunes}}{{Oliveira} et~al.}{2024}]{Fernanda24}
{Oliveira} F.,  {Avila} F.,  {Bernui} A.,  {Bonilla} A.,   {Nunes} R.~C.,  2024, \mn@doi [European Physical Journal C] {10.1140/epjc/s10052-024-12953-w}, \href {https://ui.adsabs.harvard.edu/abs/2024EPJC...84..636O} {84, 636}

\bibitem[\protect\citeauthoryear{Padmanabhan}{Padmanabhan}{1993}]{T_Padmanabhan}
Padmanabhan T.,  1993, Structure formation in the universe.
Cambridge university press

\bibitem[\protect\citeauthoryear{{Pandey}}{{Pandey}}{2010}]{Pandey10}
{Pandey} B.,  2010, \mn@doi [\mnras] {10.1111/j.1365-2966.2009.15852.x}, \href {https://ui.adsabs.harvard.edu/abs/2010MNRAS.401.2687P} {401, 2687}

\bibitem[\protect\citeauthoryear{{Papastergis}, {Giovanelli}, {Haynes}, {Rodr{\'\i}guez-Puebla}  \& {Jones}}{{Papastergis} et~al.}{2013}]{Papastergis13}
{Papastergis} E.,  {Giovanelli} R.,  {Haynes} M.~P.,  {Rodr{\'\i}guez-Puebla} A.,   {Jones} M.~G.,  2013, \mn@doi [\apj] {10.1088/0004-637X/776/1/43}, \href {https://ui.adsabs.harvard.edu/abs/2013ApJ...776...43P} {776, 43}

\bibitem[\protect\citeauthoryear{{Peebles}}{{Peebles}}{1967}]{Peebles67}
{Peebles} P.~J.~E.,  1967, \mn@doi [\apj] {10.1086/149077}, \href {https://ui.adsabs.harvard.edu/abs/1967ApJ...147..859P} {147, 859}

\bibitem[\protect\citeauthoryear{{Peebles}}{{Peebles}}{1993}]{Peebles93}
{Peebles} P.~J.~E.,  1993, {Principles of Physical Cosmology}.
Princeton University Press, \mn@doi{10.1515/9780691206721}

\bibitem[\protect\citeauthoryear{{Perivolaropoulos} \& {Skara}}{{Perivolaropoulos} \& {Skara}}{2022}]{Perivolaropoulos22}
{Perivolaropoulos} L.,  {Skara} F.,  2022, \mn@doi [\nar] {10.1016/j.newar.2022.101659}, \href {https://ui.adsabs.harvard.edu/abs/2022NewAR..9501659P} {95, 101659}

\bibitem[\protect\citeauthoryear{{Planck Collaboration} et~al.,}{{Planck Collaboration} et~al.}{2020}]{Planck20}
{Planck Collaboration} et~al., 2020, \mn@doi [\aap] {10.1051/0004-6361/201833910}, \href {https://ui.adsabs.harvard.edu/abs/2020A&A...641A...6P} {641, A6}

\bibitem[\protect\citeauthoryear{{Pons-Border{\'\i}a}, {Mart{\'\i}nez}, {Stoyan}, {Stoyan}  \& {Saar}}{{Pons-Border{\'\i}a} et~al.}{1999}]{Pons99}
{Pons-Border{\'\i}a} M.-J.,  {Mart{\'\i}nez} V.~J.,  {Stoyan} D.,  {Stoyan} H.,   {Saar} E.,  1999, \mn@doi [\apj] {10.1086/307754}, \href {https://ui.adsabs.harvard.edu/abs/1999ApJ...523..480P} {523, 480}

\bibitem[\protect\citeauthoryear{{Poulin}, {Bernal}, {Kovetz}  \& {Kamionkowski}}{{Poulin} et~al.}{2023}]{Poulin23}
{Poulin} V.,  {Bernal} J.~L.,  {Kovetz} E.~D.,   {Kamionkowski} M.,  2023, \mn@doi [\prd] {10.1103/PhysRevD.107.123538}, \href {https://ui.adsabs.harvard.edu/abs/2023PhRvD.107l3538P} {107, 123538}

\bibitem[\protect\citeauthoryear{{Qu} et~al.,}{{Qu} et~al.}{2024}]{Qu24}
{Qu} F.~J.,  et~al., 2024, \mn@doi [\apj] {10.3847/1538-4357/acfe06}, \href {https://ui.adsabs.harvard.edu/abs/2024ApJ...962..112Q} {962, 112}

\bibitem[\protect\citeauthoryear{{Reid}, {Percival}, {Eisenstein}, {Verde}, {Spergel}, {Skibba}, {Bahcall}  \& {Budavari}}{{Reid} et~al.}{2010}]{Reid}
{Reid} B.~A.,  {Percival} W.~J.,  {Eisenstein} D.~J.,  {Verde} L.,  {Spergel} D.~N.,  {Skibba} R.~A.,  {Bahcall} N.~A.,   {Budavari} I.,  2010, \mn@doi [\mnras] {10.1111/j.1365-2966.2010.16276.x}, \href {https://ui.adsabs.harvard.edu/abs/2010MNRAS.404...60R} {404, 60}

\bibitem[\protect\citeauthoryear{{Repp} \& {Szapudi}}{{Repp} \& {Szapudi}}{2020}]{Repp20}
{Repp} A.,  {Szapudi} I.,  2020, \mn@doi [\mnras] {10.1093/mnrasl/slaa139}, \href {https://ui.adsabs.harvard.edu/abs/2020MNRAS.498L.125R} {498, L125}

\bibitem[\protect\citeauthoryear{{Ribeiro}, {Bernui}  \& {Campista}}{{Ribeiro} et~al.}{2024}]{Bruno24}
{Ribeiro} B.,  {Bernui} A.,   {Campista} M.,  2024, \mn@doi [European Physical Journal C] {10.1140/epjc/s10052-024-12437-x}, \href {https://ui.adsabs.harvard.edu/abs/2024EPJC...84..114R} {84, 114}

\bibitem[\protect\citeauthoryear{{Riess} et~al.,}{{Riess} et~al.}{2022}]{Riess22}
{Riess} A.~G.,  et~al., 2022, \mn@doi [\apjl] {10.3847/2041-8213/ac5c5b}, \href {https://ui.adsabs.harvard.edu/abs/2022ApJ...934L...7R} {934, L7}

\bibitem[\protect\citeauthoryear{{Sch{\"o}neberg}, {Abell{\'a}n}, {S{\'a}nchez}, {Witte}, {Poulin}  \& {Lesgourgues}}{{Sch{\"o}neberg} et~al.}{2022}]{Schoneberg22}
{Sch{\"o}neberg} N.,  {Abell{\'a}n} G.~F.,  {S{\'a}nchez} A.~P.,  {Witte} S.~J.,  {Poulin} V.,   {Lesgourgues} J.,  2022, \mn@doi [\physrep] {10.1016/j.physrep.2022.07.001}, \href {https://ui.adsabs.harvard.edu/abs/2022PhR...984....1S} {984, 1}

\bibitem[\protect\citeauthoryear{{Tr{\"o}ster} et~al.,}{{Tr{\"o}ster} et~al.}{2020}]{Troster20}
{Tr{\"o}ster} T.,  et~al., 2020, \mn@doi [\aap] {10.1051/0004-6361/201936772}, \href {https://ui.adsabs.harvard.edu/abs/2020A&A...633L..10T} {633, L10}

\bibitem[\protect\citeauthoryear{{Trotta}}{{Trotta}}{2017}]{Trotta17}
{Trotta} R.,  2017, \mn@doi [arXiv e-prints] {10.48550/arXiv.1701.01467}, \href {https://ui.adsabs.harvard.edu/abs/2017arXiv170101467T} {p. arXiv:1701.01467}

\bibitem[\protect\citeauthoryear{{Tully} et~al.,}{{Tully} et~al.}{2023}]{Tully23}
{Tully} R.~B.,  et~al., 2023, \mn@doi [\apj] {10.3847/1538-4357/ac94d8}, \href {https://ui.adsabs.harvard.edu/abs/2023ApJ...944...94T} {944, 94}

\bibitem[\protect\citeauthoryear{{Verde}}{{Verde}}{2010}]{Verde10}
{Verde} L.,  2010, in , Lectures on Cosmology: Accelerated Expansion of the Universe.
Springer, Berlin, Heidelberg, pp 147--177

\bibitem[\protect\citeauthoryear{{Verde}, {Sch{\"o}neberg}  \& {Gil-Mar{\'\i}n}}{{Verde} et~al.}{2024}]{Verde23}
{Verde} L.,  {Sch{\"o}neberg} N.,   {Gil-Mar{\'\i}n} H.,  2024, \mn@doi [\araa] {10.1146/annurev-astro-052622-033813}, \href {https://ui.adsabs.harvard.edu/abs/2024ARA&A..62..287V} {62, 287}

\bibitem[\protect\citeauthoryear{Virtanen et~al.,}{Virtanen et~al.}{2020}]{scipy}
Virtanen P.,  et~al., 2020, \mn@doi [Nature Methods] {10.1038/s41592-019-0686-2}, \href {https://rdcu.be/b08Wh} {17, 261}

\bibitem[\protect\citeauthoryear{{de Carvalho}, {Bernui}, {Xavier}  \& {Novaes}}{{de Carvalho} et~al.}{2020}]{Edilson20}
{de Carvalho} E.,  {Bernui} A.,  {Xavier} H.~S.,   {Novaes} C.~P.,  2020, \mn@doi [\mnras] {10.1093/mnras/staa119}, \href {https://ui.adsabs.harvard.edu/abs/2020MNRAS.492.4469D} {492, 4469}

\bibitem[\protect\citeauthoryear{{de Santi} \& {Abramo}}{{de Santi} \& {Abramo}}{2022}]{Santi22}
{de Santi} N. S.~M.,  {Abramo} L.~R.,  2022, \mn@doi [\jcap] {10.1088/1475-7516/2022/09/013}, \href {https://ui.adsabs.harvard.edu/abs/2022JCAP...09..013D} {09, 013}

\makeatother
\end{thebibliography}




\appendix

\appendix

\section{Robustness test with curve fitting}
\label{appendix:curvefit}
For robustness, in this appendix, we also conducted a Curve Fitting of the data for the theoretical function (\ref{eq:power-law}), using the \textsc{scipy.optimize}\footnote{\url{https://docs.scipy.org/doc/scipy/reference/generated/scipy.optimize.curve_fit.html}.}\citep{scipy} package from the Open Source library in Python language, where we performed the best-fit analysis of the parameters $\gamma$ and $r_{0}$, in order that they minimize the $\chi^{2}$ function defined in equation~(\ref{eq:chi_sq}), where Cov$^{-1}$ is the inverse of the covariance matrix, and Cov is specified in equation~(\ref{eq:cov_matrix}). 
The uncertainties of the adjusted parameters, $\{ \sigma_i \}$, are given by the square roots of the diagonal elements of the covariance matrix, $\sigma_i = \sqrt{\text{Cov}_{ii}}$.

\begin{figure}
\begin{minipage}[b]{\linewidth}
\centering
\includegraphics[width=\textwidth]{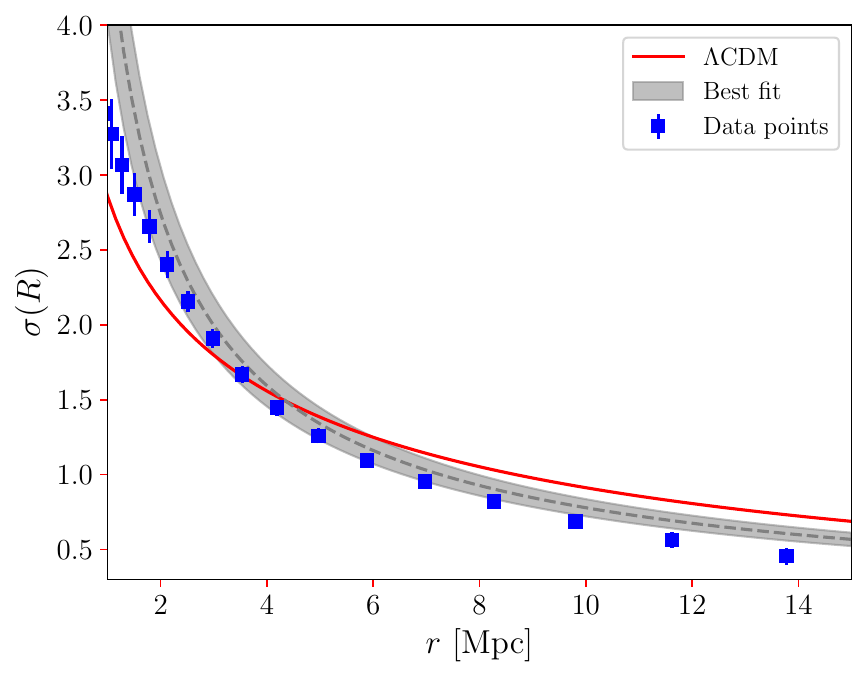}
\end{minipage}
\caption{Mass fluctuation analysis for the HI selected sample, this time using the Curve Fitting method.
}
    \label{fig:mass_fluctuation2}
\end{figure}

In figure \ref{fig:mass_fluctuation2}, the fit is shown the best fit curve, where we obtain $r_0 = 4.5 \pm 0.4 \text{; and} \, \gamma = 1.56 \pm 0.11$. As observed in table~\ref{tab:best-fits}, the adjusted parameters values are the same as those obtained using the MCMC method, with an increase in measured error bars, which is expected due to the difference in the methods employed.

\begin{table}
    \centering
    \begin{tabular}{lccc}
    \hline
        & $r_0$ [Mpc] & $\gamma$ \\
    \hline     
    \text{MCMC} & $4.5 \pm 0.3$ & $ 1.55 \pm 0.07$ \\
    \text{Curve Fitting} & $4.5 \pm 0.4$ & $ 1.56 \pm 0.11$ \\
    \hline    
    \end{tabular}
    \caption{The adjusted parameters using MCMC and Curve Fitting methods.}
    \label{tab:best-fits}
\end{table}

\section{The impact of the Virgo cluster 
as a non-linear effect}\label{appendix:virgo_cluster}
Our $\sigma_8$ measurement was obtained for an ensemble of cosmic objects with distances in the interval $20 - 85$ Mpc, a purposeful choice to avoid the overdense spatial region, where $\delta > 1$, containing the Virgo cluster. 
In this Appendix we study the impact of the presence of the Virgo cluster in our measurement. 
For this, we analyze the HI objects with distances in the interval $10 - 85$ Mpc, which includes the Virgo cluster. 
At the same time, considering for this test those cosmic objects that are closer to us we are also increasing a systematic effect sourced by large peculiar velocities~\citep{Avila23,Lopes24, Marinoni2023,Migkas2021}. 

\begin{figure}
    \begin{minipage}[b]{\linewidth}
    \centering
    \includegraphics[width=\textwidth]{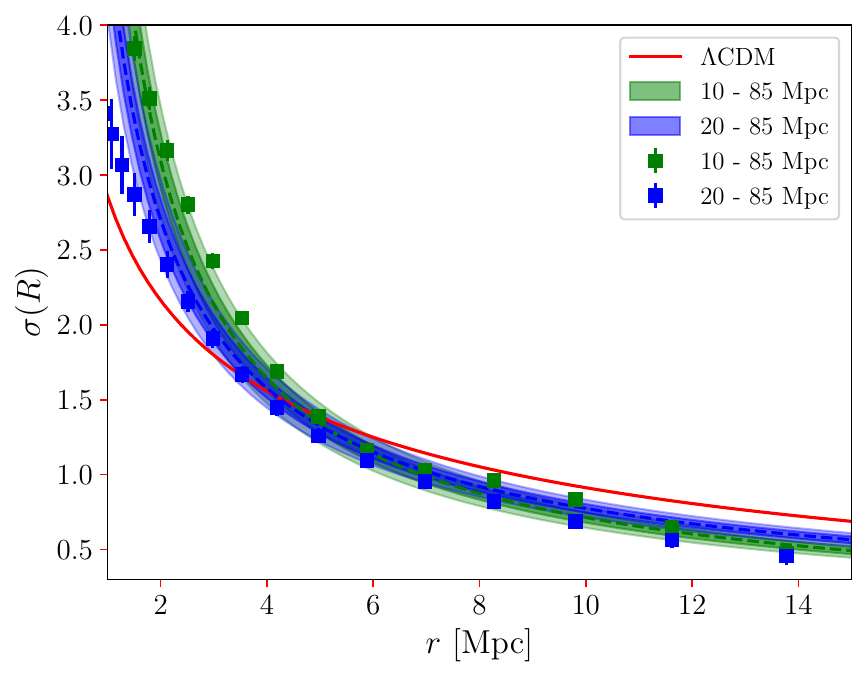}
    \end{minipage}
    \caption{
    Comparison between two ALFALFA samples 
    with HI sources belonging to distinct distance intervals. 
    The green points corresponds to the sample with sources in the interval $10-85~\text{Mpc}$; the blue points corresponds to sources in the interval $20-85~\text{Mpc}$. The shaded areas correspond to $1\sigma$ and $2\sigma$ confidence levels for each sample.
    }
    \label{fig:mass_fluctuation-comp}
\end{figure}

We perform this consistency analysis recalculating the $\sigma(R)$ function. 
Our result is shown in figure~\ref{fig:mass_fluctuation-comp}, where 
for comparison we also plot the original case. 
As expected, some differences can be observed, 
being the difference more noticeable at 
the smaller scales where the clustering, that is, the value of $\sigma(R)$, is bigger, 
presumably due to the presence of more clustered structures like Virgo. 

Nevertheless, in both cases $\sigma(R)$ agrees at the $2 \sigma$ level, confirming the robustness of our measurement.

\section{Consistency test: is the ALFALFA distance data biased with $H_0=70$ \MakeLowercase{km s}$^{-1}$ M\MakeLowercase{pc}$^{-1}$?}\label{AppendixC}

The ALFALFA collaboration provides the distance values of the HI sources, but these quantities were not directly measured but estimated assuming a semi-analytical velocity field model that uses $H_0 = 70$ km s$^{-1}$ Mpc$^{-1}$ 
for the Hubble constant~\citep{Masters2005}. 
Clearly, this hypothesis opens up the possibility that the distances released in the ALFALFA catalogue could be biased, and our $\sigma_8$ measurement too. 
To assess the potential influence of this specific $H_0$ value on our analysis, we performed a Monte Carlo (MC) procedure to generate a set of simulated catalogues of distances, termed  {\it pseudo data}, with different values of $H_0$ randomly chosen from a Gaussian distribution.

To produce $100$ MC realisations with simulated distance datasets we proceed as follows. 
For the $k$-th MC realisation, for $k=1, \cdots, 100$, the corresponding radial 
distance of the $i$-th cosmic object, $\text{\bf R}^{i}$, is obtained from 
\begin{equation}\label{eq-Ri}
\text{\bf R}^{i} = \frac{H_0}{H_{0}^{k}} \,r^{i} \,,
\end{equation}
where $H_0 = 70$ km s$^{-1}$ Mpc$^{-1}$, $r^{i}$ is the radial distance listed in the ALFALFA catalogue for the $i$-th cosmic object, and $H_{0}^{k}$ is a value randomly selected from a Gaussian distribution with mean $\overline{H_0} = 70$ km s$^{-1}$ Mpc$^{-1}$ and standard deviation $\sigma_{H_0} = 1$ km s$^{-1}$ Mpc$^{-1}$.

\begin{figure}
\begin{minipage}[b]{\linewidth}
\centering
\includegraphics[width=\textwidth]{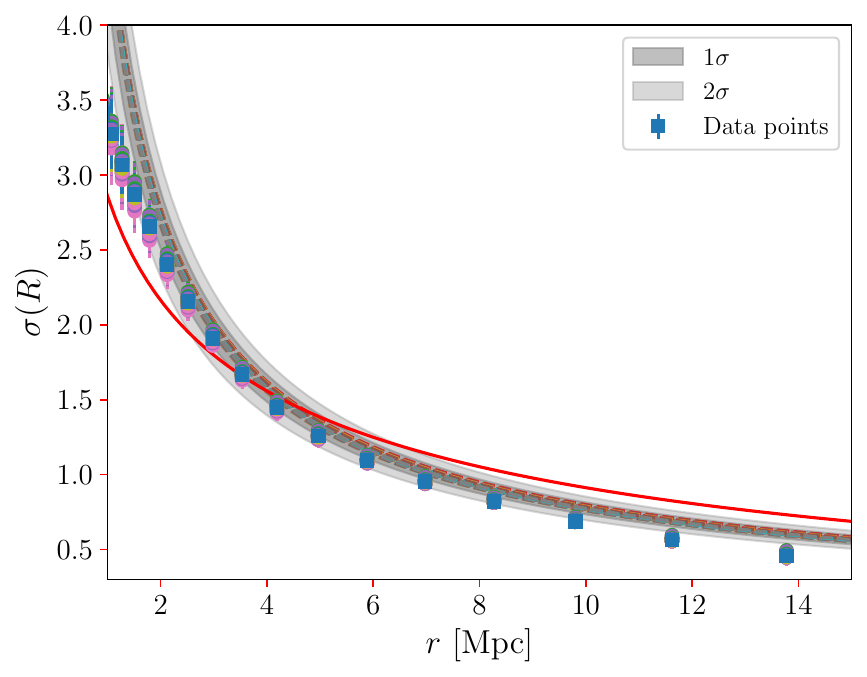}
\end{minipage}
\caption{Mass fluctuation analysis for the selected data sample from ALFALFA (blue squares), together with the analysis of the MC realisations of different distance data sets 
$\{ \text{\bf R}^{i} \}_k$, $k=1,\cdots,100$ (coloured dots). 
The blue squares with error bars and the shaded gray areas, corresponding to $1\sigma$ and $2\sigma$ confidence levels, are the same as in figure~\ref{fig:mass_fluctuation}. 
For $r \gtrsim 6$ Mpc the squares hide the dots.
}
\label{sigma_100mc}
\end{figure}

The $k$-th realisation provides the distance set 
$\{ \text{\bf R}^{i} \}_k$. 
Each one of these {\it pseudo data} sets, $\{ \text{\bf R}^{i} \}_k$, 
is analysed according to our methodology as it were the original data to calculate the function $\sigma(R)^k$ for $k = 1,\cdots, 100$. 
Finally, the set of these functions, $\{ \sigma(R)^k \}$, is then plotted in figure~\ref{sigma_100mc}, and compared with our original result shown in figure~\ref{fig:mass_fluctuation}. 
As shown by this analysis, our measurement of $\sigma_8$ is not biased with the specific value of $H_0=70$ km s$^{-1}$ Mpc$^{-1}$.

\bsp	
\label{lastpage}
\end{document}